\documentclass[aps,prd,twocolumn,nofootinbib,floatfix,showpacs,showkeys,preprintnumbers]{revtex4}
\usepackage{subfigure}
\usepackage{graphicx}
\usepackage{dcolumn}
\usepackage{epsfig}
\usepackage{amsmath}
\usepackage{amsfonts}
\usepackage{amssymb}
\usepackage{color}
\usepackage{hyperref}
\usepackage{dutchcal}

\newcommand{\be}{\begin{equation}}
\newcommand{\ee}{\end{equation}}
\newcommand{\bea}{\begin{eqnarray}}
\newcommand{\eea}{\end{eqnarray}}
\newcommand{\nn}{\nonumber}
\newcommand{\di}{\text{d}}

\begin{document}

\title{
Cosmological forecasts for future galaxy surveys with the linear point standard ruler: Toward consistent BAO analyses far from a fiducial cosmology
}

\author{Stefano Anselmi}
\email{stefano.anselmi@pd.infn.it}
\affiliation{INFN-Padova, Via Marzolo 8, I-35131 Padova -- Italy}
\affiliation{LUTH, UMR 8102 CNRS, Observatoire de Paris, PSL Research University, Universit\'e Paris Diderot, 92190 Meudon -- France}

\author{Glenn D.~Starkman}
\affiliation{Department of Physics/CERCA/Institute for the Science of Origins, Case Western Reserve University, Cleveland, OH 44106-7079 -- USA}
\affiliation{Astrophysics, Imperial College London, Blackett Laboratory, Prince Consort Road, London, SW7 2AZ -- UK}

\author{Alessandro Renzi}
\affiliation{Dipartimento di Fisica e Astronomia ``G.Galilei'', Universit\'a di Padova Via Marzolo 8, I-35131 Padova -- Italy}
\affiliation{INFN-Padova, Via Marzolo 8, I-35131 Padova -- Italy}

\date{\today}

\begin{abstract}
The linear point (LP) standard ruler was identified as the basis of a purely geometric method for exploiting the Baryon Acoustic Oscillations (BAO). The LP exploits the BAO feature imprinted in the galaxy two-point correlation function to measure cosmological distances independent of any specific cosmological model. We forecast the expected precision of future and ongoing spectroscopic galaxy surveys to measure distances leveraging the linear point. We investigate the cosmological implications of our forecasted results. We focus in particular on a relevant working example: the detection of the late-time cosmic acceleration independent of other cosmological probes. Our findings show that, even within the $\Lambda$CDM standard cosmological paradigm, estimated distances need to be reliable over a very wide parameter range in order to realize their maximum utility. This is particularly relevant if we aim to properly characterize cosmological tensions. 
The LP is a promising candidate approach to achieve this reliability.  
In contrast,
widely employed procedures in BAO analysis estimate distances keeping fixed cosmological parameters to fiducial values close to cosmic-microwave-background constraints in flat-$\Lambda$CDM. 
It is unclear whether they are purely geometric methods. 
Moreover, they rely on untested extrapolations to explore the parameter space away from those fiducial flat-$\Lambda$CDM values.
We recommend that all BAO methodologies be validated across the full range of models and parameters over which their results are quoted, first by means of linear predictions and then N-body simulations.
\end{abstract}


\maketitle

\section{Introduction}
\label{intro}

Since the first observational hints \cite{Perlmutter:1998np, Riess:1998cb} that the late time Universe is undergoing a phase of accelerated expansion, the cosmological community has invested significant effort to cross check and characterize this expansion. For this purpose, one of the most powerful acceleration probes is dubbed Baryon Acoustic Oscillations (BAO) \cite{Bassett:2009mm}. It is based on primordial Universe physics: the standard model of cosmology describes the pre-recombination Universe as a hot plasma of baryons and photons where, because of primordial inflationary perturbations, acoustic waves were generated and then propagated until decoupling. The waveform was imprinted in the distribution of baryons and, through gravitational coupling, the dark matter. A relic of those primordial waves is predicted to appear as a feature in the matter correlation function: an peak (termed the acoustic peak) at around $150$ Mpc and a dip at smaller scales. This feature is observed in the distribution of the visible tracers that track the matter field (i.e.~galaxies).

The presence of a feature in the clustering correlation function (CF) affords the possibility to map the expansion of the Universe as a function of redshift \cite{1987MNRAS.227..739S}. The feature's position can serve a {\it statistical comoving cosmological standard ruler}. This allows one to estimate the ``distance''\footnote{With ``distance'' here we refer to a variety of appropriate background quantities (e.g.~the Hubble parameter, the angular diameter distance) that are estimated from BAO analyses.} from us to the galaxy survey in units of the standard ruler length scale. This motivates the effort devoted to build large galaxy surveys such as Euclid\footnote{\url{http://sci.esa.int/euclid/}}, DESI\footnote{\url{http://desi.lbl.gov}} , 4MOST\footnote{\url{https://www.4most.eu/}}, Roman Space Telescope\footnote{\url{https://roman.gsfc.nasa.gov/}} and Subaru Prime Focus Spectrograph\footnote{\url{https://pfs.ipmu.jp/}}.

The cosmological community targeted the comoving position of the acoustic peak as a cosmological standard ruler. However, for the standard ruler to work its length needs to be redshift independent. This property was challenged by the discovery that certain late-time non-linearities distort and shift the peak position in a redshift dependent way \cite{2008PhRvD..77d3525S}. 
As shown using flat-$\Lambda$CDM numerical simulations, these distortions are significant for parameter values close to the best fit obtained from Cosmic Microwave Background (CMB) data analysis \cite{Spergel:2006hy, 2020A&A...641A...6P}. 

The BAO community proposed a fitting methodology, called {\it BAO-Only}, to overcome this problem of late-time non-linearities. Instead of using peak-finding algorithms, the full CF is fit with some phenomenological templates and some choice of nuisance parameters \cite{2008ApJ...686...13S, 2012MNRAS.427.2146X}. In doing so, the cosmological parameters are held fixed to some fiducial values in a flat-$\Lambda$CDM cosmology (even though some of these same cosmological parameters are also being estimated). This methodology was shown to be unbiased on simulations in some range of parameter values (e.g.~\cite{2020MNRAS.494.2076C}); however the arbitrary choice of the functional form of the fitting template and the parameter-fixing make error estimation for the inferred parameters ambiguous and the validity of the extension to non-$\Lambda$CDM models unclear \cite{2019PhRvD..99l3515A}\footnote{The authors of \cite{2020PhRvD.102l3515B} argue that the BAO-Only methodology holds for the standard flat-$\Lambda$CDM and for some extensions of it. While they check that the cosmological models and parameter values they considered do not significantly bias the distance measures, they ignore the error-estimation issue discussed in \cite{2019PhRvD..99l3515A}. Moreover, they only consider models and parameter values close to the Planck CMB results, in such a way as to effectively insert CMB related priors in their analysis. Finally they assume some non-linear analytic template to generate the data mocks, rather than employing N-body simulations or galaxy mocks to properly compute non-linearities.}. Other approaches to extracting cosmological information from galaxy clustering data suffer from the same problem \cite{2017MNRAS.466.2242B, 2021JCAP...12..054B}.

The standard results released by the BAO community also employ ``BAO reconstruction'' algorithms: by assuming some values for the growth rate and for the galaxy-matter bias, they approximately undo the galaxies' non-linear evolution, thereby enhancing the signal-to-noise of the BAO \cite{2007ApJ...664..675E, 2012MNRAS.427.2132P}. However, one should ask how well we can trust the resulting uncertainty propagation (i.e.~the errorbars), given that the growth rate and the matter-bias are fixed to specific fiducial flat-$\Lambda$CDM values\footnote{Other reconstruction methodologies keep fixed other parameters and/or assume knowledge of the matter field, that is unknown in real observations \cite{2021PhRvD.104d3530N, 2021MNRAS.506.1165L, 2022PhRvL.128t1302V, 2022arXiv220301868N}.}.  

To overcome problems and confusion inherent in the standard BAO methodologies, the authors of \cite{2019PhRvD..99l3515A} made a fist step to clarify what we mean by ``measuring the expansion history of the Universe'' from the BAO. They introduced the Purely-Geometric-BAO approaches (PG-BAO). These allow one to estimate distances that are geometrical (i.e. independent of the primordial fluctuation parameters over a wide range of parameters within the standard inflationary model framework) and do not assume a specific value for the spatial curvature of the Universe nor for a specific model of the late-time cosmic acceleration. In \cite{2019PhRvD..99l3515A}, two specific ways to achieve this goal were proposed. We briefly summarize them.

The first approach assumes a phenomenological CF model to properly fit the data and estimate the distance in units of the sound horizon scale $r_d$ (a secondary parameter calculated from the fitted baryon, cold-dark-matter and neutrino energy densities). Parameters that are not geometric and that depend on the late-time acceleration model are marginalized over. This approach was named correlation function model fitting (CF-MF). Crucially, given a CF model, one can correctly propagate uncertainties. Exploiting the CF-MF methodology, a preliminary Fisher-matrix investigation, found that the standard BAO-Only method underestimate the BAO distance errors by up to a factor of 2 (see Section III.C.1 of \cite{2019PhRvD..99l3515A}). 

However, since we do not know how to predict the galaxy clustering CF starting from cosmological initial conditions, a plethora of phenomenological CF models have been proposed \cite{2019PhRvD..99f3530O, 2021JCAP...11..038O} (see \cite{2022MNRAS.tmp..572A} for higher order statistics.) By choosing one of them to fit the data we assume a specific functional form, a range of scales to fit, and a set of nuisance parameters, all of which will impact the cosmological inference. While with survey mocks we can test that the CF-MF results are unbiased for some fiducial choice of parameters, how do we ensure the errorbars are properly estimated? 

The second approach precisely aims to overcome the CF-MF ambiguities. It relies on the existence of a feature in the CF that is geometrical and weakly sensitive to non-linear effects. In \cite{2016MNRAS.455.2474A} such a feature was found and dubbed {\it the linear point} (LP). It was discovered that it acts as a comoving cosmological standard ruler \cite{2016MNRAS.455.2474A, 2018PhRvL.121b1302A} and, remarkably, that its properties extend to massive neutrino cosmologies \cite{2021JCAP...01..009P}. The LP is located midway between the peak and dip positions of the clustering correlation function. Crucially, its convenient properties allows to estimate distances in a data-driven way, i.e.~without interposing a CF phenomenological model.

BAO inferred distances are supposed to be, as far as possible, independent of the cosmological model, and not to rely on informative priors on cosmological parameters. As such they are extensively used, alone or in combination with other cosmological probes, to shed light on cosmological tensions and to constrain cosmological models (e.g.~\cite{2020MNRAS.495.2630C, 2018JCAP...05..033H}). However, the reasons summarized above imply that the standard BAO results, in general, cannot be used  straightforwardly for this purpose. We need instead to use BAO distances estimated using PG-BAO methods. Moreover the inferred results need to hold for a wide parameter range. While this last point will be largely motivated through the findings of the present manuscript, it was anticipated in previous papers \cite{2019PhRvD..99l3515A, 2020PhRvD.101h3517O} and, precisely for this reason, the LP behavior was investigated over a wide parameter range \cite{2020PhRvD.101h3517O} . 

A crucial role of BAO-inferred distances is to provide strong evidence for the {\it late-time cosmic acceleration}. It is competitive with and complementary to the program of measuring supernova redshifts and luminosity distances that was used for the original persuasive detection of acceleration (e.g.~\cite{2016NatSR...635596N, 2021arXiv210812497R}). It works by assuming $\Lambda$CDM and assessing the significance of acceleration implied by the considered BAO distances.

In this manuscript, we first forecast the capability of future and ongoing spectroscopic galaxy surveys to measure cosmological distances by exploiting the linear point standard ruler. Secondly, we provide examples of how the distances can be used to learn about cosmology.  We thus employ some of the forecasted distances to detect the late-time acceleration of the Universe. For illustrative purposes we finally show how the LP measurements can be used to constrain two other  widely considered cosmologies. 

Crucially we do not aim in this work to assess the full cosmological constraining power of linear point distances. Consequently, in the cosmological investigation, we choose only a subsample of available distances. Instead, in view of the way in which BAO distances are exploited to test standard cosmology, and constrain modifications of that standard cosmology, we comment on some of the critical points inherent in BAO data-analysis pipelines\footnote{In this manuscript we focus on PG-BAO distance measurements. While some issues we discuss could be common to other approaches to clustering data-analysis (e.g.~\cite{2022arXiv220505892G}), our analysis, quantitative results and discussions are meant to be consistent within the framework of PG-BAO distance measurements.}. First, as mentioned above, the standard BAO fitting methodologies and reconstruction procedures assume specific flat-$\Lambda$CDM parameter values and CF templates. This is problematic if the measurements are then used to constrain flat-$\Lambda$CDM, and all the more so if they are used to constrain non-flat $\Lambda$CDM and Quintessence models. We remind the reader that the LP inferred distances are not affected by this inconsistency (at least within the range of models and parameters that has been established). We finally highlight that, in general, all the tools adopted to build and validate BAO analyses (e.g.~CF covariance) need to hold for a wide enough range of parameters and cosmological models. This requires, for instance, to run N-body simulations and build survey mocks in parameter ranges that have not yet been explored.

The reader should note that we make no attempt to compare the constraining power of linear point distances with those of BAO distances as traditionally employed. This is because the traditional methodologies are not consistent with the Purely-Geometric-BAO requirements. For the reader's benefit we describe specifically for each survey the difficulties in making such a comparison at this juncture.    
The reader should also note that we chose a true cosmological model and parameter values to derive the expected cosmic distances with the LP. Subsequently we employed the obtained distances to show how a cosmologist who is agnostic about the true cosmology can  constrain cosmological models. Given the obtained parameter contours, we explain that BAO distances need to be reliable over a very wide parameter range. In other words, if the BAO distance estimation procedure does not hold for all the parameter values contained in the contours those contours are not self-consistently derived. Moreover, we clarify that the BAO estimation procedure needs to be even more flexible than what we found, therefore ours is a conservative/minimum requirement. This message is independent of the choice of the true cosmological model and parameter values we made.

The manuscript is structured as follow. In Section \ref{sec:method} we explain the methodology employed: we introduce the non-linear CF model adopted; we review the linear point methodology to estimate cosmic distances; we explain how to apply the LP inferred distances to detect the late-time acceleration of the Universe; we clarify the meaning of Fisher based forecasts and detail how to adapt it to our particular investigation. In Section \ref{sec:forecasts} we introduce the spectroscopic galaxy surveys  we will consider for the linear point forecasts; we predict the expected LP inferred distance error and detection probability for each survey; we apply these results to detect the late-time cosmic acceleration. In Section \ref{sec:concl} we conclude. In the Appendix we clarify some technical details on the LP estimation procedure. Throughout our analysis we assume a standard inflationary initial
power spectrum, a standard recombination history and that the background metric is very well approximated by the Friedman-Lema\^itre-Robertson-Walker (FLRW) assumption (for non-FLRW studies see \cite{2020JCAP...01..038H}).

\section{Methodology}
\label{sec:method}

This section first presents the methodology employed to forecast the Linear Point inferred cosmic distances from future spectroscopic galaxy surveys (Sections \ref{sec:synthetic} and \ref{sec:LPdistances}). Next it explains the methodology followed  to detect the late time acceleration of the Universe and to constrain two cosmologies employing the forecasted LP distances (Sections \ref{sec:cosmology} and \ref{sec:Fisher}). In these analyses we focus on cosmological information that can be inferred without employing informative priors from other cosmological probes.

To forecast the precision and accuracy that future galaxy surveys will reach exploiting the LP standard ruler, we closely follow the methodology introduced in \cite{2019PhRvD..99l3515A} explaining in greater details the CF-mock generation and LP estimation procedures.

\subsection{Synthetic Correlation Function, covariance and CF-mocks}
\label{sec:synthetic}


\subsubsection{Correlation Function non-linear model}

In the observed redshift space the clustering correlation function is anisotropic due to redshift space distortions \cite{Kaiser:1987qv}. A convenient approach to deal with this anisotropy expands the CF in multipoles, the monopole being the one with the highest signal-to-noise. In the rest of this manuscript we will focus on the CF monopole, the relevant observable to estimate cosmological distances through the Linear Point standard ruler\footnote{Extension to the quadrupole will be the subject of future work.}. As common to many BAO data analysis setups we work in comoving fiducial coordinates.

In the BAO range of scales, the CF is affected by certain non-linear corrections, e.g. non-linear gravity, non-linearity of the real-to-redshift-space map, and scale-dependent bias related to the observed dark matter tracer (e.g.~galaxies) (see \cite{2016MNRAS.455.2474A} and reference therein). 
Following \cite{2019PhRvD..99l3515A} we describe the CF monopole through the following analytic approximation to the non-linear CF estimated from simulations:
\bea
	\xi_{0}(s) \simeq \int \frac{\di k}{k} \frac{k^{3}P_{\rm lin}(k,z)}{2 \pi^{2}} A^{2} e^{- k^{2} \sigma_{0}^{2}}\, j_{0}(ks)\,,
	\label{nl:xi}
\eea
where 
\begin{eqnarray}\label{xi:map}
	A^{2}&=&b_{10}^{2}+\frac{2 b_{10} f}{3}+\frac{f^{2}}{5}, \nn \\
	\sigma_{0}^{2}&=& \frac{\sigma_{v}^2 \left[35 b_{10}^2 \left(f^2+2 f+3\right)+14 b_{10} f \left(3 f^2+6 f+5\right)\right.}{105 \,A^{2}}  \nn \\
	&+&\quad\frac{\left.3 f^2 \left(5 f^2+10 f+7\right)\right]}{105 \,A^{2}} -\frac{2 b_{01}(3 b_{10}+f)}{3\,A^{2}}\, .
\end{eqnarray} 
Equations (\ref{nl:xi})-(\ref{xi:map}) depend on the linear power spectrum at redshift $z$, $P_{\rm lin}(k,z)$; the growth rate at redshift z, $f(z)=d \ln D/ d \ln a$; the Eulerian $b_{01}$ and scale-dependent $b_{10}$ biases; and the one-dimensional dark matter velocity dispersion computed in linear theory
\be
	\sigma_{v}^{2}(z)=\frac{1}{3} \int \frac{\di^{3} q}{(2 \pi)^{3}}\frac{ P_{\rm lin}(q,z)}{q^{2}} \, ; 
	\label{sigma:v}
\ee  
finally $j_0(x)=\sin{x}/x$ is the zero-order spherical Bessel function. 

Several considerations are in order to justify our choice of Eq.~(\ref{nl:xi}) to describe the non-linear CF.  
First, as underlined in \cite{2019PhRvD..99l3515A}, in the BAO range of scales, Eq.~(\ref{nl:xi}) closely matches the N-body simulations and galaxy mocks' CF outcomes. 
Second, even if many other CF analytical approximations have been proposed (see for instance \cite{2019PhRvD..99f3530O}), for our purposes the choice of the specific CF approximation is not particularly relevant -- it is mainly employed as a working example to apply the LP-model-independent estimation methodology. 
In this sense, in \cite{2021JCAP...01..009P} it was shown that the CF here employed can be exploited to accurately refine the LP methodology applied to CFs estimated from N-body simulations. 
Moreover, contrary to the CF-MF method to estimate cosmic distances, the LP approach itself is insensitive to the kind and number of parameters that we use to approximate the CF.

\subsubsection{Binned synthetic data, covariance and CF-mock generation}

The CF function is typically estimated in spatial bins from any specific surveyed volume. 
We thus bin the synthetic CF by band-averaging Eq.~(\ref{nl:xi}). 
In the $i$-th bin of size $\Delta{s}$ and center $s_i$ the CF is given by 
\be
	 \bar{\xi}_{0}(s_{i}) = \frac{1}{V_{s_i} } \int_{V_{s_i}} \di^{3}s \; \xi_{0}(s) \,,
	\label{binavxsi}
\ee
where $V_{s_i}$ is the volume of the spherical shell of thickness $\Delta s$
\bea
	V_{s_i}=4\pi s_i^2\Delta s \left[1+\frac{1}{12}
	\left(\frac{\Delta s}{s_i}\right)^2\right] \,.
\eea

In the BAO range of scales and in the small bin size approximation ($\Delta s/s\ll 1$) we can neglect the nonlinear covariance corrections \cite{2008PhRvD..77d3525S,2009MNRAS.400..851S,2016MNRAS.457.1577G}. We thus employ the Gaussian-Poisson approximation for the covariance of the CF among the spatial bins $i$ and $j$: 
\be
	 D_{ij} = \frac{1}{V_{\mu}} \int \frac{\di k\, k^{2}}{2 \pi^{2} } \bar{j_{0}}(k s_{i})\bar{j_{0}}(k s_{j})\sigma^{2}_{P}(k) \, .
	\label{cov}
\ee
In Eq.~(\ref{cov}) $V_{\mu}$ is the surveyed volume and $\sigma_{P}^{2}$ is defined by: 
\be
	 \sigma_{P}^{2}(k)=\int_{-1}^{1} \di \mu \left[ b_{10}^{2}(1+\beta \mu^{2})^{2}P_{\rm lin}(k) + \frac{1}{\bar{n}_{g}}  \right]^{2},
	\label{sigmaPS}
\ee
where $\bar{n}_{g}$ is the mean number density of galaxies in the survey. 
Finally we introduced $ \bar{j_{0}}$, the band-averaged zeroth-order spherical Bessel function, 
\begin{flalign} 
	\bar{{j}_0}(ks_i)&\equiv \frac{1}{\Delta s} \int_{s_{i-}}^{s_{i+}}ds {j}_0(ks_i)\,,\quad
	s_{i\pm} = s_i\pm\Delta s/2\\
	&=
	\frac{\left.s^2j_1(ks)\right|_{s_{i-}}^{s_{i+}}}
	{s_i^2k\Delta s\left[1+\frac{1}{12}\left(\frac{\Delta s}{s_i}\right)^2\right]} \,,
\end{flalign}
where $j_1(x)$ is the first-order spherical Bessel function. 

To generate the synthetic CF-monopole noisy data, we follow the methodology already implemented in \cite{2019PhRvD..99l3515A, 2021JCAP...01..009P}. This is based on \cite{2018PhRvD..98b3527A}, where it was found that the CF distribution from official BAO galaxy mocks produced by the BOSS collaboration, is always consistent with a Gaussian. Therefore, to generate the CF mocks, we assume the CF distribution is a Multivariate Gaussian: the ensemble average given by Eq.~(\ref{nl:xi}) band-averaged through Eq.~(\ref{binavxsi}); the covariance by Eq.~(\ref{cov}); both evaluated at the effective redshift of the survey, chosen here to be the mean redshift of the selected redshift bin.

\subsection{Cosmic distances with the Linear Point}
\label{sec:LPdistances}

\subsubsection{Isotropic-volume-distance from a correlation function feature}

As stated above, we work in comoving coordinates; however, in the real Universe, we measure angles and redshifts, not comoving coordinates. We thus need to interpose a fiducial cosmological model to translate the measured quantities into fiducial comoving coordinates (see \cite{2013MNRAS.434.2008S,2011MNRAS.411..277S,2019MNRAS.487.3419M} for alternative approaches) . Hence our observables are affected by the so-called Alcock-Paczynski distortion effect. In particular the estimated CF will be distorted w.r.t.~the true CF \cite{2013MNRAS.431.2834X}. 

The  CF monopole  Alcock-Paczynski distortions are conveniently described in terms of the isotropic-volume-distance
\be
	D_{V}(z) \equiv \left[(1+z)^{2}D_{A}(z)^{2}\frac{cz}{H(z)}\right]^{{1/3}}\, ,
	\label{xi:y}
\ee
where $H(z)$ is the Hubble rate and $D_A(z)$ is the angular-diameter distance.

The distorted CF is related to the undistorted one by \cite{2013MNRAS.431.2834X}
\be
	\tilde{\xi}_{0}^{{\rm fid}}(y^{{\rm fid}}) \simeq \tilde{\xi}_{0}^{{\rm true}}(y^{{\rm true}})\, ,
\label{xianyy}
\ee
where we labelled as ``fid'' and ``true'' the distorted and the true correlation function respectively. 
In Eq.~(\ref{xianyy}) we have introduced $y^{{\rm x}}(\bar{z}) \equiv s^{{\rm x}}/D_V^{{\rm x}}(\bar{z})$ and the reduced correlation function (RCF) $\tilde{\xi}_{0}^{{\rm x}} (t)\equiv \xi_{0}^{{\rm x}} ( D_{V}^{{\rm x}}(\bar{z}) t)$, with ${{\rm x}}=\{{{\rm true}},\, {{\rm fid}}\}$.

On BAO scales, corrections to Eq.~(\ref{xianyy}) are negligible provided that the fiducial values of the Universe's energy densities are sufficiently close to the true ones (see e.g.~\cite{2016MNRAS.457.1770C}). 

As shown in \cite{2019PhRvD..99l3515A} the functions $\tilde{\xi}_{0}^{{\rm fid}}$ and $\tilde{\xi}_{0}^{{\rm true}}$ are equal. Therefore, if $y_1^{{\rm fid}}$ is the location of a feature of the fiducial RCF, it is also the location of the true RCF, namely $y_1^{{\rm true}}$. It follows 
\be  
\frac{s_1^{{\rm true}}}{D_V^{{\rm true}}}
\simeq
\frac{s_1^{{\rm fid}}}{D_V^{{\rm fid}}}\, .
\label{ygeom}
\ee
Thanks to Eq.~(\ref{ygeom}) and given a particular CF feature, we will be able to estimate $s_1^{{\rm true}}/D_V^{{\rm true}}(z)$, loosely called ``cosmic distance''. 

Henceforth, exploiting Eq.~(\ref{ygeom}) and to ease the reading of the manuscript, we shall drop the superscripts ``fid'' and ``true''.

\subsubsection{Linear Point model-independent estimation}
\label{sec:LPest}

Eq.~(\ref{ygeom}) can be exploited to provide a geometrical estimate of cosmic distances if we find a correlation-function feature that is a cosmological standard ruler, i.e.~is independent of the primordial cosmological parameters and of the redshift of the survey. This is the case for the Linear Point --- it is geometrical and weakly sensitive to redshift-independent late-time nonlinearities. Moreover, the LP position is insensitive to the Dark Energy density and to the specific value of the spatial curvature of the Universe.

Notably, the LP feature can be estimated in a data-driven way, i.e.~without choosing a cosmology-dependent non-linear CF template. 
In practice, in \cite{2018PhRvL.121b1302A, 2018PhRvD..98b3527A} it was proposed to exploit a simple polynomial interpolation. 
We first fit the galaxy RCF data with a polynomial function:
\be
\label{eqn:polynomialfit}
\xi_{0}^{\rm fit}(y)=\sum_{i=0}^{n}a_i y^i \, .
\ee 
(See \cite{2018PhRvD..98b3527A, 2021JCAP...01..009P} for detailed validation of the polynomial estimator.)
Recalling that the LP is defined by
\be
	y_{LP}=\frac{1}{2}(y_{\rm peak}+y_{\rm dip}),
\ee
its estimate from data $\hat{y}_{\rm LP}$ is obtained by computing the numerical solutions of $d\xi_{0}^{\rm fit}/dy=0$ to find the dip ($\hat{y}_{\rm dip}$) and peak ($\hat{y}_{\rm peak}$) positions. 
The error in the LP is estimated by first Taylor expanding $\hat{y}_{\rm LP}$ w.r.t.~the polynomial coefficients and then applying linear error-propagation \cite{2021JCAP...01..009P}.

Following \cite{2016MNRAS.455.2474A}, we finally multiply $\hat{y}_{\rm LP}$ by $1.005$ to obtain an estimate consistent with the linear theory LP at the 0.5\% level. In this regard a comment is in order. The LP was tested against several numerical experiments. Initially \cite{2016MNRAS.455.2474A}, the LP was characterized by means of dark matter N-body simulations. These showed that the LP was weakly affected by non-linear corrections, in real and redshift space, for both the dark matter field and halos. 
In \cite{2021JCAP...01..009P, 2021PhRvD.104d3530N}, the same small secular shift was confirmed ($\sim 1\%$), while in \cite{2021PhRvD.104f3504N, 2022arXiv220301868N} a slightly larger correction was found ($\sim 1.5\%$). It is unclear whether this difference depends on the different halo mass considered or on the assumed recipes to populate dark matter halos with galaxies, or it might simply be due to a spurious disagreement among different dark matter N-body simulations. Such a spurious disagreement among simulations was found in \cite{2021JCAP...01..009P} (see discussion related to their Fig.~4) and \cite{2021PhRvD.104d3530N} (see their Appendix C). We advocate that this accuracy limit in predictions from N-body simulations  (i.e.~the reference theoretical tools) be properly accounted for and reported in clustering investigations.

\subsubsection{Linear Point estimated from the CF-mocks and its detection probability} 
\label{DetProb}

Given the CF-mocks, we estimate the LP following the methodology described in section \ref{sec:LPest}.
As explained in greater detail in \cite{2018PhRvD..98b3527A}, the LP estimation procedure can be conveniently optimized. 
The main steps involved can be summerized as follows. 

For each redshift survey and bin we generate, following Section \ref{sec:synthetic}, 1000 synthetic CF realizations with the optimal binning choice of 3 Mpc/h \cite{2018PhRvD..98b3527A}.  
We then fit the polynomial \eqref{eqn:polynomialfit} to each CF mock, assuming a Gaussian likelihood, consistent with the discussion and procedure explained in Section \ref{sec:synthetic}.

Given the finite volume of the surveys and the finite number of galaxies, i.e.~cosmic and sample variance, the CF BAO feature might not be present. Operatively, this means that the polynomial estimator previously defined would not ``detect'' the peak and dip in the BAO range of scales, i.e.~$d\xi_{0}^{\rm fit}/dy=0$ would not have real solutions. 
In each redshift survey and bin we count the number of realizations for which there is no LP detection. 
We estimate in this way the probability that a future galaxy survey will detect the LP.  

We first check that the distribution of $\chi_{\rm min}^2$, obtained from fitting the polynomial to the CF mocks, is consistent with the expected $\chi^2$ distribution,
and that the mean of the estimated LP is unbiased w.r.t.~the true value computed from Eq.~(\ref{nl:xi}). 
By varying the CF range-of-scale over which to perform the polynomial fit,
we maximize the LP detection probability while minimizing the LP statistical error. 
The whole procedure returns us with the optimal range-of-scales.
We have found a quintic polynomial to always provide enough degrees of freedom to properly fit the synthetic data\footnote{Notice that for CFs estimated from high resolution simulations a higher order polynomial might be needed \cite{2021JCAP...01..009P}.}.

In Appendix \ref{appendix:LPestimation} we comment on more detailed aspects of the analysis.


\subsection{Detecting the Late Time Acceleration of the Universe and cosmological constraints}
\label{sec:cosmology}

In cosmology, arguably the most important application of ``Purely-Geometric-BAO cosmological distances'' is to detect the late-time acceleration of the Universe without relying on prior information from other cosmological probes. 
As carefully explained in \cite{2019PhRvD..99l3515A}, the LP standard ruler allows to estimate $s_{LP}/D_V(z)$ without assuming that the spatial curvature is flat or that the late-time acceleration of the Universe is driven by a cosmological constant. 
Moreover, the LP estimation procedure does not need to assume a cosmological CF model template, 
i.e.~$y_{LP}(z)$ depends only on the cosmological model and not on the additional theoretical assumptions employed to model the clustering 2-point correlation function.

Within the class of cosmological models encompassed by the PG-BAO distances, we can write the LP estimated quantity at redshift $z$, making explicit its parameter dependences \cite{2016MNRAS.455.2474A, 2020PhRvD.101h3517O, 2021JCAP...01..009P}
\be
	y_{LP}(z)=\frac{s_{LP}(\omega_b,\, \omega_c,\, \omega_{\nu},\omega_\gamma)}{D_V(z;\, H_0,\, \Omega_m,\, \Omega_K,\, \Omega_{DE}, \, w(z))}\, .
	\label{dist:param}
\ee
Here $\omega_b,\, \omega_c,\, \omega_{\nu}$, and $\omega_\gamma$ represent the current energy densities of baryons, cold dark matter, neutrinos, and photons respectively. 
($\Omega_i$ are ratios of current energy densities to the critical energy density $8\pi G/3H_0^2$.)
$\Omega_m \equiv \Omega_b + \Omega_c + \Omega_{\nu}$ (since, at late times, neutrinos scale as matter), $H_0$ is the Hubble constant, $\Omega_K$ is the energy density currently associated with the spatial curvature, $\Omega_{DE}=1-\Omega_m-\Omega_K$ represents the current Dark Energy density and $w(z)$ is the Dark Energy equation-of-state parameter (EOS).
($\omega_\gamma$ is very well measured, and small compared to the density of other cosmological constituents today; we fix it in $s_{LP}$
and  neglect it in $D_V$.)

In summary, to constrain cosmological quantities, we can assume a cosmological model allowed by the LP data-analysis and properly combine $y_{LP}$ measurements performed by different galaxy surveys, each one properly subdivided in redshift bins. 

In the following we show how to exploit the LP estimated distances to constrain cosmology. We start with the detection of the late-time acceleration and the cosmological constant. We then utilize the LP distances to constrain two widely adopted cosmologies where the extra assumption of spatial flatness is added; even though this assumption may not be theoretically well-motivated it is considered, by the majority of the cosmological community, an essential ingredient of the standard model of cosmology; moreover, it is always employed in almost all the validation processes built for BAO data-analysis. We therefore also show how to apply the distance measurements to constrain cosmological models that assume spatial flatness.  Similarly, for simplicity and given the absence of a preferred Dark Energy model other than a cosmological constant ($w=-1$), we consider a constant EOS, even though there is no compelling theoretical motivation for constant $w\neq-1$. For data-analysis with real clustering data, either a theoretically well-motivated DE model that predicts a specific $w(z)$ should be employed, or one should utilize a data-driven approach to $w(z)$ like principal component analysis \cite{2003PhRvL..90c1301H}.

Given the above discussion and choices, at low redshift we can write the Friedman equation as: 
\be
	H(z)^{2}=H_{0}^2[\Omega_{m}(1+z)^{3}+\Omega_{k}(1+z)^{2}+\Omega_{DE}(1+z)^{3 (1+w)}]\, ,
	\label{Hubble}
\ee  
where $w$ is a constant.

The angular diameter distance is given by
\be
	D_{A}(z)=\frac{1}{(1+z)}S\left( r(z) \right)\, ,
	\label{da}
\ee  
where
\be
	 r(z)\equiv  c \int_{0}^{z} \frac{\text{d} z'}{H(z')}\, ,
\ee
$c$ is the speed of light and
\bea
	S(r) \equiv 
	\begin{cases} 
		\frac{\sin(r \sqrt{-\Omega_{K}}H_{0} /c )}{ \sqrt{-\Omega_{K}}H_{0}/c} & \Omega_{K}<0,\\ 
		r  & \Omega_{K}=0,\\ 
		\frac{\sinh(r \sqrt{\Omega_{K}}H_{0} /c )}{ \sqrt{\Omega_{K}}H_{0}/c} & \Omega_{K}>0. \\ 
	\end{cases}
\eea

As written above we consider the following three cosmological models:
\begin{enumerate}
\item $\Lambda$CDM: in Eq.~(\ref{Hubble}) the EOS is fixed to $w=-1$;
\item flat-$\Lambda$CDM: in Eq.~(\ref{Hubble}) the EOS is fixed to $w=-1$ and the curvature is fixed to $\Omega_{k}=0$;
\item flat-$w$CDM: in Eq.~(\ref{Hubble}) the curvature is fixed to $\Omega_{k}=0$.
\end{enumerate}

\subsubsection{Detection of the late-time acceleration and cosmological constant}
\label{sec:late-timeAcc}

We are interested in the ability of future surveys to detect the late-time acceleration of the Universe within the $\Lambda$CDM model using the PG-BAO. This implies that\footnote{We require the deceleration of the Universe to be negative 
\be
	 q(z)\equiv  -a \frac{\ddot{a}}{a^{2}} = -1 -\frac{\dot{H}}{H^{2}}\, .
\ee
}
to high significance $\Omega_{m}\le 2\,\, \Omega_{\Lambda}$.
(We recall that in this model $\Omega_{DE}\equiv\Omega_{\Lambda}$).

Since, in the $D_{V}$ definition, $H_{0}$ and the speed of light factorize out, it is convenient to define

\be
	d_{V}(z;\, \Omega_{m}, \Omega_{\Lambda}) \equiv D_{V}(z;\, H_{0}, \Omega_{m}, \Omega_{\Lambda})\,H_{0}/c\, 
	\label{dv}
\ee
Recalling that the comoving size of the LP is independent of the primordial cosmological parameters, for each redshift bin, the measured quantity reads
\be
	\frac{s_{LP}}{D_{V}(z)} = \frac {H_{0}\, s_{LP}(\omega_b,\, \omega_c,\, \omega_{\nu})}{c}\, \frac{1}{d_{V}(z;\, \Omega_{m}, \Omega_{\Lambda})}\, .
\ee
Exploiting the standard-ruler properties of the LP, we treat  $c/[H_{0}\, s_{LP}(\omega_b,\, \omega_c,\, \omega_{\nu})]$ as a single redshift-independent parameter.

\subsubsection{flat-$\Lambda$CDM}
If we assume that the spatial curvature of the Universe is flat, Eq.~(\ref{dv}) becomes
\be
	d_{V}^{f}(z;\,\Omega_{m}) \equiv D_{V}(z;\, H_{0}, \Omega_{m})\, H_{0}/c\, . 
	\label{dvf}
\ee
Therefore the parameter dependence of $y_{LP}(z)$ is
\be
	\frac{s_{LP}}{D_{V}(z)} = \frac {H_{0}\, s_{LP}(\omega_b,\, \omega_c,\, \omega_{\nu})}{c}\, \frac{1}{d_{V}^f(z;\, \Omega_{m})}\, .
\ee
Again $c/[H_{0}\, s_{LP}(\omega_b,\, \omega_c,\, \omega_{\nu})]$ is a redshift-independent parameter that is marginalized over.

\subsubsection{Flat-wCDM}

Within the flat-$w$CDM model, we are interested in constraining both the dark energy EOS parameter $w$ and $\Omega_{m}$. 
It is again useful to make the parameter dependence explicit and adopt the following definition
\be
	d_{V}^{w}(z;\,\Omega_{m}, w) \equiv D_{V}(z;\, H_{0}, \Omega_{m}, w)\, H_{0} / c\, . 
	\label{dvw}
\ee

We are interested in the LP measured quantity:
\be
	\frac{s_{LP}}{D_{V}(z)} = \frac {H_{0}\, s_{LP}(\omega_b,\, \omega_c,\, \omega_{\nu})}{c}\, \frac{1}{d_{V}^w(z;\, \Omega_{m}, w)}\, .
\ee
The combination $c/[H_{0}\, s_{LP}(\omega_b,\, \omega_c,\, \omega_{\nu})]$ is marginalized over.

\subsection{Forecast Methodology}
\label{sec:Fisher}

In this section, we present the procedure we follow to forecast the ability of future galaxy surveys to constrain the cosmological models presented in section \ref{sec:cosmology} given the LP distance measurements. The idea is to select non-overlapping redshift bins from different surveys, estimate $\hat{y}_{LP}(z)$ for each bin, combine all the bins, and compare them with the predictions of the cosmological models. Since in Appendix \ref{appendix:LPestimation} we find the LP best-fit distributions to be consistent with a Gaussian distribution, to describe the probability that the measurements are drawn from a specific cosmological model, we  assume a Gaussian likelihood for the data. Consistent with the purely-geometric-BAO approach, we assume that there is no correlation between the distance measures of different redshift bins.

As argued in \cite{1997ApJ...480...22T}, the inverse of the Fisher matrix can be used as an estimate of the covariance matrix,
and so, as is usually done, we employ the Fisher matrix to characterize the likelihood function around best fit parameter values.

Given a set of mocks of the LP distance measurements  $\{s_{LP}/D_V(z_1),... ,s_{LP}/D_V(z_n)\}$ in $n$ pre-selected redshift bins with effective redshifts $z_i$, we are interested in forecasting the errors on the cosmological parameters. 
These ``$y_{LP}$ mocks'' are generated with specific values for the cosmological parameters. We will show how the uncertainty in the bias of the LP (around 0.5\%) translates into uncertainty in the biases of the  cosmological parameters. It is only those uncertainties that limit the ability to correct for the induced biases.

Given a set of parameters $p_\alpha$, the Fisher information matrix is\footnote{Summation over repeated indices is implied and $,\!\alpha \equiv \partial_{p_\alpha}$.}
\be
	F_{\alpha \beta} \equiv \langle (\log L)_{,\alpha} (\log L)_{,\beta}  \rangle \, , 
\ee
where $L$ is the likelihood function. For Gaussian data
\be
	L =  \frac{1}{(2 \pi)^{n/2} \sqrt{|C|} } e^{-\frac{1}{2}X_i C^{-1}_{ij} X_j}\, ,
\ee
where $X_i = m_i - \mu_i$, with $m_i$ representing the data in the $i$-th redshift bin and $\mu_i$ the theoretical prediction of the model for that bin; $C_{i,j}$ is the covariance matrix between the bins (self-consistently assumed to be diagonal as explained above). 
Given that the purely-geometric-BAO methods estimate distances in cosmology-model-independent ways, the covariance matrix will be considered to be independent of cosmological parameters. This should be confirmed with dedicated studies (see Section \ref{sec:ResultsAcc} below).

Given the above, the Fisher matrix can be conveniently written as
\bea
	\label{eq:fisher}
	F_{\alpha \beta} 
				 &=& \frac{1}{4} \langle  [(\mu_{i,\alpha}X_j+\mu_{j,\alpha}X_i) C^{-1}_{ij}]\,\\
				 &&\qquad\qquad [(\mu_{i,\beta}X_j+\mu_{j,\beta}X_i) C^{-1}_{ij}]  \rangle\, . \nn 
\eea
Usually, to ease the computation, Eq.~(\ref{eq:fisher}) can be greatly simplified (e.g.~\cite{1997ApJ...480...22T}). 
However, we are confronted with the special case where $<m_i > \neq  \mu_i$, because of the uncertainty in the bias of the LP. 
Moreover, and perhaps more importantly, as explained in section \ref{DetProb}, the probability of detecting the LP is not always $100\%$ for every redshift bin. 
We therefore cannot rely on the calculation proposed in \cite{1997ApJ...480...22T}. 

We estimate $F_{\alpha \beta}$ using \eqref{eq:fisher} and 1000 $y_{LP}$ mocks -- i.e. 1000 realizations of ${\bf m}$ (see section \ref{DetProb}). 
We have already learned from the earlier study of the CF-mocks (see above) the probability of a false negative, i.e.~no LP being detected, for each redshift bin of any given $y_{LP}$-mock. 
Therefore the number of redshift bins in which there is a measurement can change from $y_{LP}$-mock to $y_{LP}$-mock. 
The covariance matrix $C_{ij}$ is estimated from the distributions of $y_{LP}(z)$ from the CF-mocks.

The Fisher matrix computation does not provide information on the bias in the cosmological parameters introduced by the LP uncertainty in the bias. To assess that we rely on the usual parameter-bias definition
\be
\label{parbias}
	b_{p_\alpha} =  <\hat{p}_\alpha> - p^{\rm true}_\alpha\, ,
\ee
where $\hat{p}_\alpha$ is the best-fit value of $p_\alpha$ obtained from a single realization of ${\bf m}$ and $p^{\rm true}_\alpha$ is the true value of the parameter. We estimate $<\hat{p}_\alpha>$ by means of our 1000 realizations of ${\bf m}$. In order to improve the correspondence between the parameter covariance matrix and the inverse of the Fisher information matrix, in  Eq.~(\ref{eq:fisher}) we will evaluate $\mu_i$ and its derivatives not on $p^{\rm true}_\alpha$ but on $ <\hat{p}_\alpha>$, i.e.~we force the best-fit to be unbiased. In this way we expect to mitigate the spurious impact of the uncertainty in the bias on the parameter-covariance-matrix estimation.

Notice that the numerical values of the Fisher matrix depend on the chosen fiducial cosmology parameters. This is not a drawback for our investigation while it needs to be taken into account in Bayesian model selection studies \cite{2006MNRAS.369.1725M}.

\section{Forecasts for Stage-IV galaxy surveys}
\label{sec:forecasts}

In this section we perform forecasts for relevant future and ongoing galaxy surveys. 
We first briefly summarize the expected details of the surveys. 
We then present the forecasted errors on the distance measures $s_{LP}/D_{V}(\bar{z})$. 
We finally employ the forecasted results to show how we can constrain the late-time acceleration of the Universe and the cosmological models outlined in section \ref{sec:cosmology}.

\subsection{Fiducial Model Parameters and Galaxy Survey Characteristics}
\label{sec:surveys}

Throught the following numerical investigation, we assume a fiducial flat-$\Lambda$CDM model with cosmological parameter values close to the best-fit results found by the Planck team, adopting values from \cite{2016A&A...594A..13P}: $\Omega_{b} = 0.0486$, $\Omega_{c}= 0.259$, $H_{0} = 67.74$, $n_{s}=0.9667$ and $\sigma_{8}=0.831$.  

We examine the following spectroscopic galaxy surveys.
\begin{itemize} 

\item {\bf Euclid} \\
We consider a Euclid-like survey that mimics the expected performance of the ESA Euclid satellite. We employ the redshift-bin definitions, number density of observed $H_\alpha$ galaxies, sky fraction $f_{{\rm sky}}=0.364$, and linear galaxy-matter bias reported in Tables 2 and 3 of \cite{2020A&A...642A.191E}.

\item {\bf DESI} \\
The DESI survey is designed according to \cite{2016arXiv161100036D}. The target galaxies are Luminous Red Galaxies (LRGs), Emission Line Galaxies (ELGs) Bright Galaxies (BGs) and quasi-stellar objects (QSOs). Following \cite{2016arXiv161100036D}, we assume a sky fraction  $f_{{\rm sky}}=0.339$. The expected number density of galaxies per unit redshift per square degree is reported in Tables 2.3 and 2.5 of  \cite{2016arXiv161100036D}. The redshift dependence of the linear galaxy-matter bias is specified in terms of the matter growth factor $D(z)$ normalized to $D(z=0)\equiv 1$. For the four selected DESI targets:
\bea
	&&b^{LRG}_{10}(z)\, D(z)=1.7 \nn \\
	&&b^{ELG}_{10}(z)\, D(z)=0.84  \nn \\
	&&b^{BGS}_{10}(z)\, D(z)=1.34 \nn \\
	&&b^{QSO}_{10}(z)\, D(z)=1.2\, .
\eea

\item {\bf 4MOST} \\
We model the 4MOST galaxy survey following \cite{2019Msngr.175...50R}. The 4MOST survey will measure cosmological distances targeting LRGs, BGs and QSOs. The sky fraction employed for the BAO survey is $f_{{\rm sky}}=0.18$. Table 1 of \cite{2019Msngr.175...50R} reports the forecast number of galaxies and relative redshift ranges, which we adopt. We employ the linear galaxy-mass bias assumed for the DESI survey. 

\item {\bf Nancy Grace Roman Space Telescope (Roman)} \\
We consider the Nancy Grace Roman Space Telescope High Latitude Spectroscopic Survey. 
We consider the reference design presented in \cite{2021arXiv211001829W}, with a sky fraction $f_{{\rm sky}}=0.0485$ and optimistic dust attenuation. 
The target galaxies will be measured in the $H_{\alpha}$ and $[O_{III}]$ bands. 

\item {\bf Subaru Prime Focus Spectrograph (PFS)} \\
The Subaru Prime Focus Spectrograph is a spectroscopic survey. It will target $[O_{II}]$ ELGs. 
We model the PFS  following \cite{2014PASJ...66R...1T}. The sky fraction is $f_{{\rm sky}}=0.0355$.
\end{itemize}

Since there is no forecast on the value of the scale-dependent bias for the considered galaxy surveys, we set it to zero. 
For all the surveys we neglect errors in spectroscopic redshifts.

\begin{table}
\vspace{1cm}
\renewcommand{\arraystretch}{1.1}
\begin{center}
\textbf{Euclid-like survey} \\

{\scriptsize
\begin{tabular}{l c c c c c c c c c c c c}\\
\hline \\
$\bar{z}$ &&& $\Delta z$ &&&&&  $\sigma_{s_{LP}/D_{V}(\bar{z})}$   &&& LP detection probability \\ [0.3cm]
\hline 
\hline \\ 
$1.0$        &&&       $0.2$    &&&&&     $1.0\%$    &&&    $100\%$     \\ [0.3cm]
$1.2$        &&&       $0.2$    &&&&&     $1.0\%$    &&&    $100\%$     \\ [0.3cm]
$1.4$        &&&       $0.2$    &&&&&     $1.1\%$    &&&    $100\%$     \\ [0.3cm]
$1.65$        &&&       $0.3$    &&&&&     $1.0\%$    &&&    $100\%$     \\ [0.3cm]
\hline
\end{tabular}
}
\end{center}
\
\caption[]{\label{tab:Eucl} 
We show the expected percentual error on the distance measurements obtained with the LP standard ruler from a Euclid-like survey. The probability of the LP detection is also shown. }
\end{table}

\begin{table}
\vspace{1cm}
\renewcommand{\arraystretch}{1.1}
\begin{center}
\textbf{DESI} \\
{\scriptsize
\begin{tabular}{l c c c c c c c c c c c c c c c c}\\
\hline \\
\multicolumn{3}{c}{} &&&&& \multicolumn{3}{c}{\small BGS} &&&&& \multicolumn{3}{c}{\small ELG} \\
\\
\hline \\
$\bar{z}$ &&  $\Delta z$ &&&&&  $\sigma_{s_{LP}/D_{V}(\bar{z})}$  && LP-det. &&&&&  $\sigma_{s_{LP}/D_{V}(\bar{z})}$  && LP-det. \\ [0.3cm]
\hline 
\hline \\ 
$0.15$    &&    $0.3$    &&&&&     $3.1\%$    &&    $71\%$     &&&&&     $-$    &&    $-$  \\ [0.3cm]
$0.4$    &&    $0.2$    &&&&&     $2.4\%$    &&    $87\%$     &&&&&     $-$    &&    $-$  \\ [0.3cm]
$0.75$    &&    $0.3$    &&&&&     $-$    &&    $-$     &&&&&     $1.2\%$    &&    $100\%$  \\ [0.3cm]
$1.0$    &&    $0.2$    &&&&&     $-$    &&    $-$     &&&&&     $1.1\%$    &&    $100\%$  \\ [0.3cm]
$1.2$    &&    $0.2$    &&&&&     $-$    &&    $-$     &&&&&     $1.1\%$    &&    $100\%$  \\ [0.3cm]
$1.4$    &&    $0.2$    &&&&&     $-$    &&    $-$     &&&&&     $1.7\%$    &&    $100\%$  \\ [0.3cm]
$1.6$    &&    $0.2$    &&&&&     $-$    &&    $-$     &&&&&     $2.6\%$    &&    $96\%$  \\ [0.3cm]
\vspace{0.3cm} \\
\hline \\
\multicolumn{3}{c}{} &&&&& \multicolumn{3}{c}{\small LRG} &&&&& \multicolumn{3}{c}{\small QSO} \\
\\
\hline \\
$\bar{z}$ &&  $\Delta z$ &&&&&  $\sigma_{s_{LP}/D_{V}(\bar{z})}$  && LP-det. &&&&&  $\sigma_{s_{LP}/D_{V}(\bar{z})}$  && LP-det. \\ [0.3cm]
\hline 
\hline \\ 
$0.75$    &&    $0.3$    &&&&&     $0.9\%$    &&    $100\%$     &&&&&     $4.5\%$    &&    $86\%$  \\ [0.3cm]
$1.05$    &&    $0.3$    &&&&&     $1.9\%$    &&    $100\%$     &&&&&     $4.2\%$    &&    $92\%$  \\ [0.3cm]
$1.35$    &&    $0.3$    &&&&&     $-$    &&    $-$     &&&&&     $3.7\%$    &&    $96\%$  \\ [0.3cm]
$1.6$    &&    $0.2$    &&&&&     $-$    &&    $-$     &&&&&     $3.8\%$    &&    $96\%$  \\ [0.3cm]
$1.8$    &&    $0.2$    &&&&&     $-$    &&    $-$     &&&&&     $3.6\%$    &&    $93\%$  \\ [0.3cm]
\hline
\end{tabular}
}
\end{center}
\
\caption[]{\label{tab:DESI} 
We show the expected percentual error on the distance measurements obtained with the LP standard ruler from the DESI survey. The probability of the LP detection is also shown. }
\end{table}

\begin{table}
\vspace{1cm}
\renewcommand{\arraystretch}{1.1}
\begin{center}
\textbf{4MOST} \\
{\scriptsize
\begin{tabular}{l c c c c c c c c c c c c c c c c}\\
\hline \\
\multicolumn{3}{c}{} &&&&& \multicolumn{3}{c}{\small BGS} &&&&& \multicolumn{3}{c}{\small LRG} \\
\\
\hline \\
$\bar{z}$ &&  $\Delta z$ &&&&&  $\sigma_{s_{LP}/D_{V}(\bar{z})}$  && LP-det. &&&&&  $\sigma_{s_{LP}/D_{V}(\bar{z})}$  && LP-det. \\ [0.3cm]
\hline 
\hline \\ 
$0.25$    &&    $0.25$    &&&&&     $3.2\%$    &&    $77\%$     &&&&&     $-$    &&    $-$  \\ [0.3cm]
$0.55$    &&    $0.3$    &&&&&     $-$    &&    $-$     &&&&&     $1.5\%$    &&    $97\%$  \\ [0.3cm]
\vspace{0.3cm} \\
\hline \\
\multicolumn{3}{c}{} &&&&& \multicolumn{3}{c}{\small QSO} &&&&& \multicolumn{3}{c}{\small } \\
\\
\hline \\
$\bar{z}$ &&  $\Delta z$ &&&&&  $\sigma_{s_{LP}/D_{V}(\bar{z})}$  && LP-det. &&&&&    &&  \\ [0.3cm]
\hline 
\hline \\ 
$1.05$    &&    $0.3$    &&&&&     $3.7\%$    &&    $87\%$     &&&&&     $$    &&    $$  \\ [0.3cm]
$1.35$    &&    $0.3$    &&&&&     $4.0\%$    &&    $92\%$     &&&&&     $$    &&    $$  \\ [0.3cm]
$1.65$    &&    $0.3$    &&&&&     $4.0\%$    &&    $94\%$     &&&&&     $$    &&    $$  \\ [0.3cm]
$2.0$    &&    $0.4$    &&&&&     $3.3\%$    &&    $96\%$     &&&&&     $$    &&    $$  \\ [0.3cm]
\hline
\end{tabular}
}
\end{center}
\
\caption[]{\label{tab:4MOST} 
We show the expected percentual error on the distance measurements obtained with the LP standard ruler from the 4MOST survey. The probability of the LP detection is also shown.}
\end{table}

\begin{table}
\vspace{1cm}
\renewcommand{\arraystretch}{1.1}
\begin{center}
\textbf{Nancy Grace Roman Space Telescope} \\
{\scriptsize
\begin{tabular}{l c c c c c c c c c c c c c c c c}\\
\hline \\
\multicolumn{3}{c}{} &&&&& \multicolumn{3}{c}{\small $H_{\alpha}$} &&&&& \multicolumn{3}{c}{\small $[O_{III}]$ } \\
\\
\hline \\
$\bar{z}$ &&  $\Delta z$ &&&&&  $\sigma_{s_{LP}/D_{V}(\bar{z})}$  && LP-det. &&&&&  $\sigma_{s_{LP}/D_{V}(\bar{z})}$  && LP-det. \\ [0.3cm]
\hline 
\hline \\ 
$1.1$    &&    $0.2$    &&&&&     $1.8\%$    &&    $97\%$     &&&&&     $-$    &&    $-$  \\ [0.3cm]
$1.3$    &&    $0.2$    &&&&&     $1.5\%$    &&    $99\%$     &&&&&     $-$    &&    $-$  \\ [0.3cm]
$1.5$    &&    $0.2$    &&&&&     $1.5\%$    &&    $100\%$     &&&&&     $-$    &&    $-$  \\ [0.3cm]
$1.7$    &&    $0.2$    &&&&&     $1.5\%$    &&    $100\%$     &&&&&     $-$    &&    $-$  \\ [0.3cm]
$1.9$    &&    $0.2$    &&&&&     $1.9\%$    &&    $99\%$     &&&&&     $-$    &&    $-$  \\ [0.3cm]
$2.1$    &&    $0.2$    &&&&&     $-$    &&    $-$     &&&&&     $3.3\%$    &&    $97\%$  \\ [0.3cm]
$2.3$    &&    $0.2$    &&&&&     $-$    &&    $-$     &&&&&     $2.8\%$    &&    $96\%$  \\ [0.3cm]
$2.55$    &&    $0.3$    &&&&&     $-$    &&    $-$     &&&&&     $3.6\%$    &&    $97\%$  \\ [0.3cm]
$2.85$    &&    $0.3$    &&&&&     $-$    &&    $-$     &&&&&     $4.2\%$    &&    $92\%$  \\ [0.3cm]
\hline
\end{tabular}
}
\end{center}
\
\caption[]{\label{tab:rom} 
We show the expected percentual error on the distance measurements obtained with the LP standard ruler from the Roman galaxy survey. The probability of the LP detection is also shown. }
\end{table}

\begin{table}
\vspace{1cm}
\renewcommand{\arraystretch}{1.1}
\begin{center}
\textbf{Subaru Prime Focus Spectrograph (PFS)} \\

{\scriptsize
\begin{tabular}{l c c c c c c c c c c c c}\\
\hline \\
$\bar{z}$ &&& $\Delta z$ &&&&&  $\sigma_{s_{LP}/D_{V}(\bar{z})}$   &&& LP detection probability \\ [0.3cm]
\hline 
\hline \\ 
$0.7$        &&&       $0.2$    &&&&&     $5.4\%$    &&&    $76\%$     \\ [0.3cm]
$0.9$        &&&       $0.2$    &&&&&     $3.5\%$    &&&    $88\%$     \\ [0.3cm]
$1.1$        &&&       $0.2$    &&&&&     $2.9\%$    &&&    $92\%$     \\ [0.3cm]
$1.3$        &&&       $0.2$    &&&&&     $2.4\%$    &&&    $96\%$     \\ [0.3cm]
$1.5$        &&&       $0.2$    &&&&&     $2.5\%$    &&&    $97\%$     \\ [0.3cm]
$1.8$        &&&       $0.4$    &&&&&     $2.0\%$    &&&    $99\%$     \\ [0.3cm]
$2.2$        &&&       $0.4$    &&&&&     $2.1\%$    &&&    $100\%$     \\ [0.3cm]
\hline
\end{tabular}
}
\end{center}
\
\caption[]{\label{tab:PFS} 
We show the expected percentual error on the distance measurements obtained with the LP standard ruler from the PFS survey. The probability of the LP detection is also shown. }
\end{table}

\subsection{Forecasted cosmic distances with the Linear Point}

We divide each galaxy surveys introduced in section \ref{sec:surveys} into narrow redshift bins. This allows us to obtain tomographic redshift information, crucial to constraining the dynamics of dark energy.
In practice, in order to estimate the distance-measurement errors, we consider redshift bins that are small enough to neglect the cosmological evolution within the bin but large enough (i.e. $\Delta z \geq 0.2$) to detect the BAO in the longitudinal direction. 
As is usually done in forecast studies, we assume the effective redshift of the bin corresponds to the mean redshift.

In Table \ref{tab:Eucl}, we present the distance-error estimates, i.e.~the expected error for $y_{LP}=s_{LP}/D_V(\bar{z})$, for a Euclid-like survey. In addition we report, for each redshift bin, the probability of detecting the LP.  
Notice that our LP forecasts cannot be directly compared with the official Euclid cosmological forecast results presented in \cite{2020A&A...642A.191E}.  
That study does not report the values of the PG-BAO distance measurements nor the choice of priors  adopted for the fit  and marginalized parameters.

In Table \ref{tab:DESI}, we report the error on the LP distance estimates obtained for the DESI survey. 
For several reasons,
it is not possible to directly compare the LP distance measurements with those presented in \cite{2016arXiv161100036D}.
In \cite{2016arXiv161100036D},  PG-BAO methods are not employed for CF-MF. 
For example, the error propagation does not comply with the PG-BAO requirements (see \cite{2019PhRvD..99l3515A}). Moreover, it is assumed that the CFs are reconstructed in the BAO range of scales; i.e.~using a procedure that requires one to assume values for unknown parameters. 
The distance measurements reported in \cite{2016arXiv161100036D} combine different galaxy populations when they overlap, while we conservatively assume we do not know the cross-correlation coefficient. 
Finally, a strong CMB prior is assumed to assess the precision of the sound horizon scale, while we eschew any informative prior from other experiments.

In Table \ref{tab:4MOST}, the expected errors relative to LP-inferred distances are shown for the 4MOST survey. 
The LP detection probability is also reported for each redshift bin. 
In \cite{2019Msngr.175...50R}, detailed forecasts of BAO-inferred distances are not presented, 
therefore we cannot compare with our results.

Table \ref{tab:rom} reports the LP distance error and detection probability forecast with the expected design of the Roman telescope. 
In \cite{2021arXiv211001829W}, an approach similar to the PG-BAO CF-MF \cite{2019PhRvD..99l3515A} is used to forecast the BAO distance measures. 
However, the non-linear parameters are kept fixed and not marginalized over; 
moreover, it is not clear which priors are assumed to compute the Fisher matrix. 
We recall that both these steps are crucial to properly estimate errors for PG-BAO distances \cite{2019PhRvD..99l3515A, 2020PhRvD.101h3517O}. 
In conclusion, the distance measures presented in \cite{2021arXiv211001829W} are not derived with full PG-BAO methods, 
hence comparing them with the LP distances would be misleading. \\
 
The LP distance errors and detection probability expected for the PFS survey are presented in Table \ref{tab:PFS}. 
The official PFS forecast paper \cite{2014PASJ...66R...1T} does not employ the CF-MF procedure needed to infer PG-BAO measurements \cite{2019PhRvD..99l3515A}. 
For example, the error propagation does not comply with the PG-BAO requirements and the BAO observables are assumed to be reconstructed; as written above with reference to DESI, this is not consistent with the PG-BAO approaches to inferring cosmic distances. 
Also, contrary to the PG-BAO working conditions, in \cite{2014PASJ...66R...1T} CMB cosmological information is assumed in the Fisher matrix computation. 
These crucial differences w.r.t.~the PG-BAO procedure likely explain why the LP expected results are slightly less constraining than the forecast  errors found in the official PFS forecast manuscript.

Another reason, 
common to all the presented surveys, that prevents us from comparing the LP distances with the official forecast distance measures, is that the LP uses an MCMC approach to estimate distances while the official survey forecasts employ the Fisher matrix approach, despite the known risk of underestimating the errors. 
Probably less significantly,  contrary to the LP analysis presented here, the official forecasts take no account of the possibility of non-detection of the BAO in some redshift bins of a survey.  While this is likely to be less of an issue with their approaches than with the LP, it is left unquantified.

The official papers tend to employ the (anisotropic) power spectrum rather than the CF as the BAO observable.  The relative merits, and biases, of each is not fully understood. Given that both contain the same information one might well prefer to observe the agreement of inference methods employing each observable.  
Finally, currently, the LP PG-BAO approach employs only the CF monopole. We expect that incorporating the CF quadrupole, the subject of current study, will reduce forecast LP error bars if it can be done in a PG-BAO manner.

We remind the reader that some of the Euclid and DESI linear point forecasts were already presented in \cite{2019PhRvD..99l3515A}. For both surveys in \cite{2019PhRvD..99l3515A}, the LP detection probability was not considered. Moreover, the Euclid forecasts are now updated w.r.t~\cite{2019PhRvD..99l3515A} to match the new instrument and survey specifications and new observations of $H_\alpha$ galaxy number densities \cite{2020A&A...642A.191E}. For DESI, in \cite{2019PhRvD..99l3515A} the LP distance errors for BGS were slightly underestimated because the linear galaxy-bias formula valid for Euclid $H_\alpha$ galaxies was erroneously employed. This small discrepancy was not relevant for the results of \cite{2019PhRvD..99l3515A}; the CF-MF and  LP computations both assumed the same bias values and in \cite{2019PhRvD..99l3515A} the relevant message was about a comparison between the CF-MF and the LP cosmic distances.

We note that the LP detection probability will be higher when we deal with real data analysis. One can adjust the redshift bin boundaries in order to maximize the number of bins where the LP is detected, hopefully allowing the LP to be detected in all redshift bins.

To conclude, in this subsection we presented, for relevant spectroscopic galaxy surveys, the forecast error and detection probability of the cosmic distances estimated by means of the LP standard ruler. The LP forecasts here presented are the first available that belong to the class of Purely-Geometric-BAO distance measurements, a promising candidate method to infer cosmic distances over a wide range of $\Lambda$CDM cosmological parameters relatively far from the Planck best-fit, and for a certain cosmological models other than flat-$\Lambda$CDM.

\begin{table}
\vspace{1cm}
\renewcommand{\arraystretch}{1.1}
\begin{center}
\textbf{Illustrative choice of redshift bins} \\

{\scriptsize
\begin{tabular}{l c c c c c c c c c c c c}\\
\hline \\
$\bar{z}$ &&& $\Delta z$ &&&&&  $\sigma_{s_{LP}/D_{V}(\bar{z})}$   &&& LP detection probability \\ [0.3cm]
\hline 
\hline \\ 
$0.15$        &&&       $0.3$    &&&&&     $3.1\%$    &&&    $71\%$      \\ [0.3cm]
$0.4$        &&&       $0.2$    &&&&&     $2.4\%$    &&&    $87\%$      \\ [0.3cm]
$0.75$        &&&       $0.3$    &&&&&     $0.9\%$    &&&    $100\%$     \\ [0.3cm]
$1.0$        &&&       $0.2$    &&&&&     $1.0\%$    &&&    $100\%$     \\ [0.3cm]
$1.2$        &&&       $0.2$    &&&&&     $1.0\%$    &&&    $100\%$      \\ [0.3cm]
$1.4$        &&&       $0.2$    &&&&&     $1.1\%$    &&&    $100\%$     \\ [0.3cm]
$1.65$        &&&       $0.3$    &&&&&     $1.0\%$    &&&    $100\%$     \\ [0.3cm]
$1.9$        &&&       $0.2$    &&&&&     $1.9\%$    &&&    $99\%$    \\ [0.3cm]
 $2.2$        &&&       $0.4$    &&&&&     $2.1\%$    &&&    $100\%$   \\ [0.3cm]
$2.55$        &&&       $0.3$    &&&&&     $3.6\%$    &&&    $97\%$      \\ [0.3cm]
$2.85$        &&&       $0.3$    &&&&&     $4.2\%$    &&&    $92\%$   \\ [0.3cm]
\hline
\end{tabular}
}
\end{center}
\
\caption[]{\label{tab:selected} 
Expected percentual error and detection probability obtained throught the LP standard ruler. We show a selection of redshift bins taken from the considered galaxy surveys. 
The selection is not intended to provide the largest constraining power, but to show and discuss relevant cosmological examples of distances measured throught a PG-BAO method.}
\end{table}

\subsection{Constraining Cosmology}
\label{sec:confidence}

In this section, we showcase the numerical results obtained exploiting the LP distance measures to constrain cosmology. 
We select the cosmological frameworks introduced in Section \ref{sec:cosmology} and apply the methodology detailed in Section \ref{sec:Fisher} to forecast the expected errors on the cosmological parameters.

We first select non-overlapping redshift bins of different galaxy surveys that are expected to measure $y_{LP}$ with small errors. As noted above, this is just an illustrative choice that does not include the full constraining power encoded in the forecast distances for the considered surveys\footnote{Therefore we do not have to worry about calculating the value of the covariance between different galaxy-surveys. It would have been easy to compute for certain cases, such as 4MOST and DESI whcih observe complimentary parts of the sky, but more involved in most other cases.}. Forecasting the full constraining power is not the goal of this investigation.  

The selected bins, the values of $y_{LP}$ and the LP detection probability are shown in Table \ref{tab:selected}.

\begin{figure}
\centering
\includegraphics[width=1\hsize]{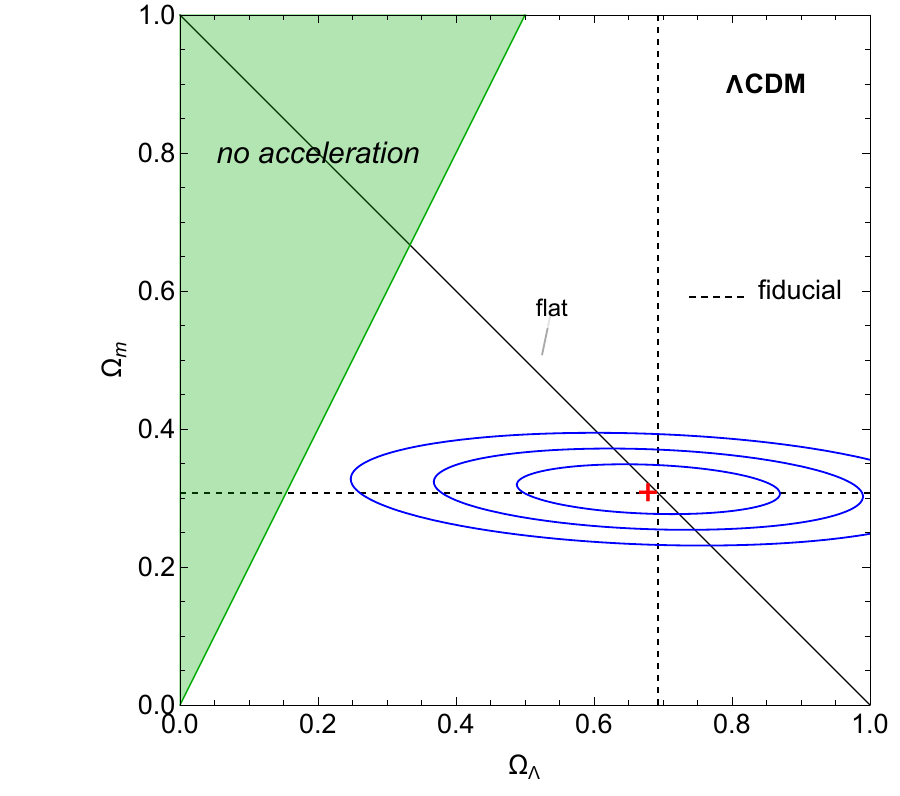}
\caption{\label{fig:contAcc} 
Constraining the late-time acceleration of the Universe employing the selected {\em linear point} cosmic distances presented in Table \ref{tab:selected}. The red ``+'' symbol is the center of the ellipses, it corresponds to the expectation values of the parameters. The bias w.r.t.~the fiducial parameter values is given by the LP uncertainty in the bias. We remind that we do not include the full constraining power of the considered surveys (see text). Therefore we expect in the future to be able, with the linear point, to constrain the late-time acceleration of the Universe with a higher significance than the $3-4\sigma$ obtained here.}
\end{figure}

\begin{figure}
\centering
\includegraphics[width=1\hsize]{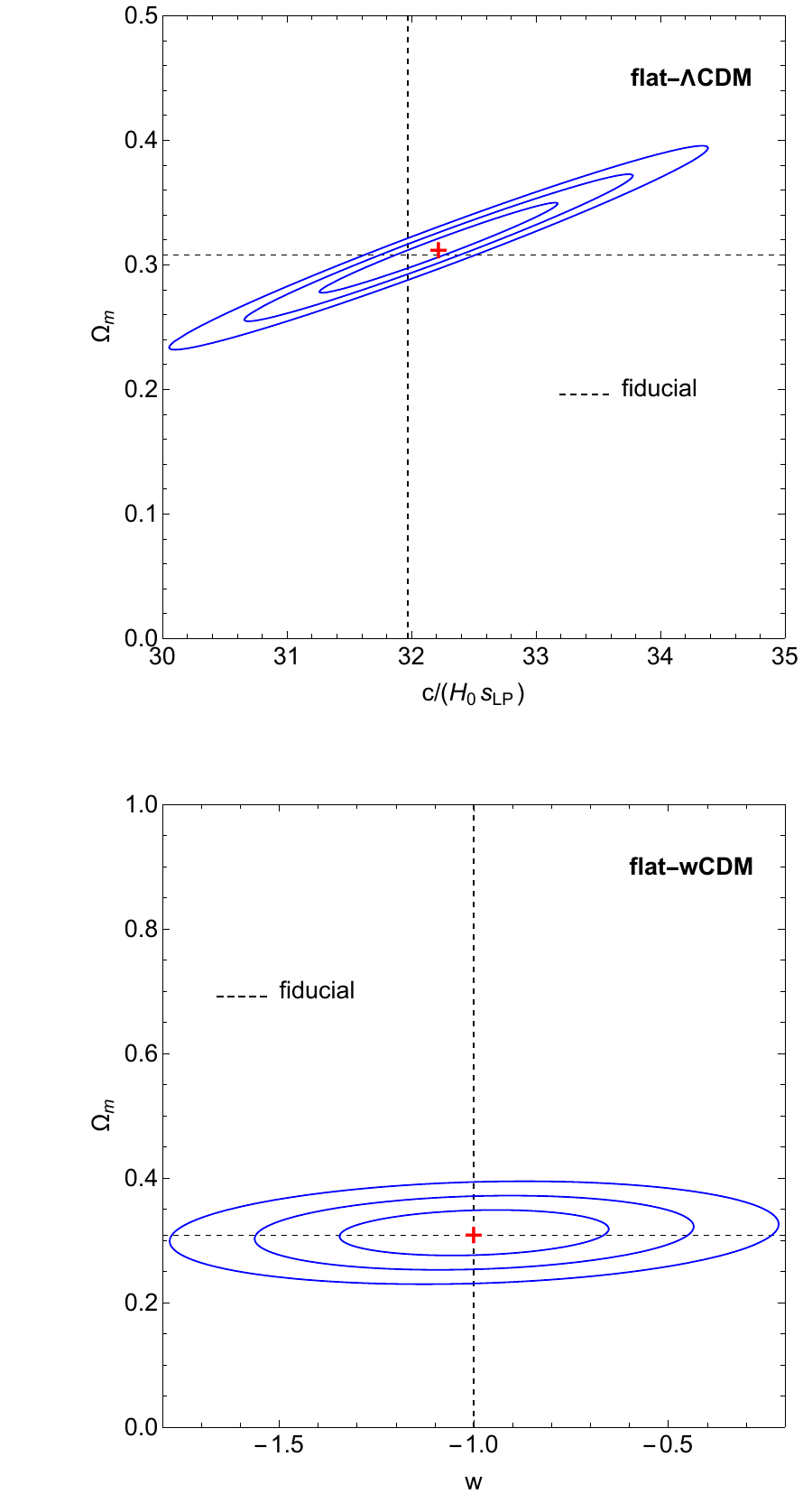}
\caption{\label{fig:contFlat} 
Constraining  flat-$\Lambda$CDM and the Dark Energy equation-of-state parameter employing the selected {\em linear point} cosmic distances presented in Table \ref{tab:selected}. The red ``+'' symbol is the center of the ellipses, it corresponds to the expectation values of the parameters. The bias w.r.t.~the fiducial parameter values is given by the LP uncertainty in the bias. As for Fig.~\ref{fig:contAcc} we remind that we do not include the full constraining power of the considered surveys.}
\end{figure}

\subsubsection{Detecting the late time acceleration of the Universe}
\label{sec:ResultsAcc}

In Fig.~\ref{fig:contAcc}, we show the estimated 1-2-3$\sigma$ intervals obtained with the selected redshift bins employed to constrain the late-time acceleration of the Universe as described in Section \ref{sec:cosmology}. (See Table \ref{tab:selected}.) 
To show the impact of the LP uncertainty in the bias we center the ellipses on $<\hat{p}_\alpha>$, where $p_\alpha$ is the set of chosen parameters and $\hat{p}_\alpha$ is its best-fit estimate. The red ``+'' symbol is the center of the ellipses. The dashed lines indicate the fiducial values of the parameters.  Note that the fiducial parameter values are well within the $1\sigma$ contour.  

The green colour in Fig.~\ref{fig:contAcc} shows the non-accelerating region of parameter space, as explained in Section \ref{sec:late-timeAcc}. Notice that, given our selection of forecast distances, the exclusion of current and past galaxy survey results, and the absence of the CF quadrupole information, we expect in the future to be able to constrain the late-time acceleration of the Universe with a higher significance than the $3-4\sigma$ we read from Fig.~\ref{fig:contAcc}. Another way to calculate the cosmic acceleration would be to project this two-dimensional parameter space onto the deceleration parameter $q$, compute the PDF of $q$, and measure the detection significance of $q<0$; however, we resist doing so here because $q$ is not usually regarded as a fundamental parameter of $\Lambda$CDM. 

Results similar to Fig.~\ref{fig:contAcc} are considered by the cosmological community to be one of the main probes of the late-time cosmic acceleration. In this respect, the most constraining probes are BAO (e.g.~\cite{2013MNRAS.436.1674A, 2015PhRvD..92l3516A, 2020PhRvL.124v1301N, 2021PhRvD.103h3533A})\footnote{Even if the $\Omega_\Lambda$ detection is often confused with the detection of the late-time Universe acceleration \cite{2020PhRvL.124v1301N}, that is instead properly defined in Section \ref{sec:cosmology}.} and Type Ia SN  (e.g.~\cite{2016NatSR...635596N, 2021arXiv210812497R}). However, regarding the BAO results, we recall that the distance measurements customarily employed were obtained by building the whole data-analysis pipeline with flat-$\Lambda$CDM survey mocks and N-body simulations close to the Planck best-fit cosmology. Moreover, standard-BAO distances are obtained through fitting approaches where cosmological parameters were kept fixed to a fiducial flat-$\Lambda$CDM cosmological model (see Section \ref{intro}). Published plots like Fig.~\ref{fig:contAcc} are not, as one would want, the result of a MCMC procedure where, for each point in the MCMC, the relevant cosmological and non-cosmological parameters are varied. Those plots are rather obtained with an unverified extrapolation procedure, and may be completely misleading. \\
Furthermore, the standard-BAO distances usually employed are obtained through BAO reconstruction, thus attempting to reconstruct the shape of the linear CF by means of a procedure that, instead of varying, keeps fixed the fiducial value of the growth rate and of the linear galaxy bias in a flat-$\Lambda$CDM model close to the Planck CMB best fit one. The reconstruction procedure does not allow to interpret the obtained measures far from a narrow prior close to the fiducial flat-$\Lambda$CDM model.  

We advocate making the BAO detection of the late-time acceleration of the Universe credible over a wide range of $\Lambda$CDM parameters by employing a consistent treatment of the data.
Even more challenging, one would like every single-redshift BAO-distance to be applicable to DE models and non-flat cosmologies. This implies that the region ideally described by BAO-distances is even wider than what we show in Fig.~\ref{fig:contAcc}. 
Hence BAO-distances need to be either stress-tested or employed with extreme caution to properly infer cosmological information.
We plan to test the BAO data-analysis pipeline for a wide parameter range running appropriate N-body simulations and building galaxy-mocks.\\

A step forward to ameliorate the standard-BAO late-time Universe acceleration analysis would employ the CF-MF PG-BAO method; however, as mentioned above, it is subject to ambiguities w.r.t.~the chosen CF non-linear model. In addition, the available analytical CF non-linear models have not been tested in a wide enough parameter range. 

The BAO distances inferred with the LP standard ruler (i.e.~a PG-BAO method) are not affected by the ambiguity problem of the CF-MF. Moreover a preliminary investigation was performed to ensure that, in the linear approximation, the results derived with the LP are sensible for a wide parameter range \cite{2020PhRvD.101h3517O}. We advocate that, when employing the BAO as a proof of the late-time Universe acceleration, the analysis presented in \cite{2020PhRvD.101h3517O} needs to be extended by taking into account the non-linear correction and by populating N-body simulations with different galaxy prescriptions.

It is also important to test the covariance dependence of the PG-BAO results. The covariance employed in the fitting procedure is always fixed to the fiducial cosmology parameter values. It is assumed that its parameter dependence has a negligible impact on the final cosmological inference. However, given the wide parameter range that is required to detect the late-time acceleration of the Universe, this assumption should be questioned and tested. In this respect, the cosmological parameters, the models and parameters used to populate halos with galaxies need to be varied. Investigations in this direction for the matter power spectrum found non-negligible results \cite{2021MNRAS.500.2532B}.

The above mentioned data-analysis consistency problems have not yet been addressed, even with the simplifying assumption that the linear approximation is the correct description for effects such as gravitational dynamics, redshift-space-distortions and the galaxy-matter bias relation. This preliminary investigation will be subject of future work.

Finally, constraints like those of Fig.~\ref{fig:contAcc} allow for $\Omega_K$ values significantly different from zero. However the CF Alcock-Paczynski distortion equations used by BAO studies (e.g.~\cite{2013MNRAS.431.2834X}) are based on Euclidean geometry. The accuracy of a Euclidean approximation for large $\vert\Omega_K\vert$ should be assessed. We will address this issue in future work.

We recall that BAO are considered by the cosmological community to be one of the main probes that show the Universe is spatially flat, i.e.~$\Omega_K$ is compatible with zero. For example, in \cite{2020MNRAS.496L..91E} it was stressed that if one combines the Planck satellite results with the BAO distance measurements one finds the Universe is spatially flat. However, we stress once again that the official BAO distances are not inferred with a PG-BAO procedure, hence we should take this result with caution. In this regard, and for the purpose of combining the LP distance measurements with CMB results it is useful to show how the LP uncertainty on the bias translates in a bias in $\Omega_K$. We found $<\hat{\Omega}_K>=+0.0071$  (recall that $\Omega^{\rm true}_K=0$). Hence there is a small bias towards negatively curved Universe (given our assumption of a flat-$\Lambda$CDM fiducial model). In fact, in our example the uncombined LP distances gives $\sigma_{\Omega_K}=0.12$, a value much larger than $b_{\Omega_K}$. However, when CMB and BAO are combined, the statistical error can be greatly reduced; moreover it is unpredictable how the $b_{\Omega_K}$ we found impacts the result of the data combination. Nevertheless it is reassuring to know that the LP distances will not spuriously drive $\Omega_K$ in the negative direction, i.e.~the one selected by the CMB data \cite{2020A&A...641A...6P}. Therefore, should the combination of LP-BAO and CMB gives a positively curved Universe, this will unlikely be caused by the LP uncertainty in the bias.

\subsubsection{Constraining flat-$\Lambda$CDM and the Dark Energy equation of state parameter}

We employ the PG-BAO distances presented in Table \ref{tab:selected} to constrain the flat-$\Lambda$CDM and the flat-$w$CDM cosmological paradigms. The results are shown in Fig.~\ref{fig:contFlat}. The 1-2-3$\sigma$ intervals are centred, as described above about Fig.~\ref{fig:contAcc}, on the parameter mean values. The dashed lines indicate the fiducial values of the parameters. For both cosmologies, the LP uncertainty in the bias gives results within the $1\sigma$ contour.

It can be read off from the plots that the $\Omega_m$ constraining power is the same for all the three cosmologies considered, i.e.~the marginalized $\Omega_m$ error is $\simeq 8\%$. This is probably due to the weak correlation of $\Omega_m$ with both the $\Omega_K$ and $w$ parameters.

We found that the bias of the parameters (i.e.~Eq.~(\ref{parbias})) is very close for all the three cosmological models considered. These means that the parameter extensions to flat-$\Lambda$CDM (i.e.~$\Lambda$CDM and  flat-$w$CDM) shift just slightly the average of the likelihood best fit w.r.t.~flat-$\Lambda$CDM.

Finally it is worth underlying that the LP uncertainty in the bias causes a negligible bias on the recovered value of $w$. In fact $b_{w} \sim 0.2\%$, better than very optimistic forecasts on our ability to measure the DE equation of state parameter. 

We remind the reader that the detail criticisms and possible solutions, presented in Section \ref{sec:ResultsAcc}, to standard-BAO approaches apply also to plots similar to Fig.~\ref{fig:contFlat} that one could find in literature.

\section{Conclusions}
\label{sec:concl}
Despite the great success of $\Lambda$CDM, we are currently facing a situation where different experiments have statistically significant disagreements on the values of cosmological parameters, such as $H_0$ and $\sigma_8$ \cite{2021arXiv210505208P}.  
Certain experiments  also display statistically anomalous features, such as the absence of large-angle correlations in the CMB \cite{2016CQGra..33r4001S}.
We are thus concerned about the self-consistency of the underlying model.
In light of these tensions, cosmologists need to revisit the historical Bayesian approach of always combining different data-sets to reduce statistical uncertainties on inferred model parameters. 
We also need to  properly quantify tensions among cosmological probes. 

At the level of the prior choice in parameter determination or model selection, information is typically properly treated (e.g.~\cite{2014A&A...571A..16P}) in cosmological Bayesian analysis;  however, 
unquantified and unclear prior information may enter in more subtle ways.
This is the case, for instance, with traditional approaches to the Baryon Acoustic Oscillations feature in the clustering correlation function and power spectrum.
The standard BAO methodology to estimate cosmological distances from galaxy surveys, which involves both BAO reconstruction and fitting of a model template,
 injects unquantified cosmological prior information in the data-analysis \cite{2019PhRvD..99l3515A}. 
Throughout, cosmological parameters are kept fixed to fiducial values in flat-$\Lambda$CDM, even though some of these same parameters (or functions of them) are also being contextually estimated.

The Purely-Geometric-BAO \cite{2019PhRvD..99l3515A} requirements are a set of necessary conditions to avoid the undesired injection of cosmological information into distances inferred from the BAO.
Traditional methods do not satisfy these requirements. As a consequence a preliminary Fisher-based analysis found they underestimate the error bars \cite{2019PhRvD..99l3515A}.
In contrast, the {\it linear point} (LP) standard ruler does satisfy the requirements; it is a powerful tool \cite{2016MNRAS.455.2474A, 2019PhRvD..99l3515A} for measuring cosmological distances from the BAO without assuming a template model for the clustering correlation function \cite{2018PhRvL.121b1302A}. 
It is therefore a method by which cosmological results from BAO can be consistently compared to those from other cosmological probes.

In this paper, we have forecast the expected error in cosmic distances inferred by means of the LP standard ruler. 
We considered future and ongoing galaxy spectroscopic surveys, namely a Euclid-like survey and the DESI, 4MOST, Roman Space Telescope and Subaru Prime Focus Spectrograph surveys. 
Because our purpose is to use the BAO to compare with other cosmological probes, 
these error bars have been forecast without the injection of prior information from those other probes,
i.e. using the LP, a Purely Geometric BAO technique.
We should therefore not expect the resulting error bars on distances to necessarily be competitive with non-PG-BAO techniques, which do
incorporate additional information in an uncontrolled unquantified way.
Nevertheless, for each survey, we explained why the forecast errors we found cannot be compared to the official forecasts for that survey.
In addition to the issue of information injection,  we find that many such forecasts are based on a Fisher matrix methodology, which is known to often underestimate statistical errors, a problem that is not present in our MCMC based analysis. 

In addition to inferring cosmological distances, we would like to use those distances to make inferences on cosmological parameters.  
Thus we wish to forecast error bars on those cosmological parameters.
In this regard, to demonstrate the utility of the LP as a PG-BAO technique for arriving at truly independent estimates of $\Lambda$CDM parameters, or of searching for extensions of $\Lambda$CDM, we have performed a proof-of-concept demonstration of the LP as a stand-alone probe of late-universe acceleration. 
We show how a selection of  forecasted LP-inferred distances can be used to detect the current cosmic acceleration. 
Our selection is purposely far from optimal: the BAO distances are intended to be flexible enough to be used in many different ways, allowing one to perform multiple consistency checks of data and theory \cite{2020MNRAS.495.2630C}. 
Such analyses are particularly relevant in light of the current cosmological tensions. 
Our selection and usage of distances therefore does not attempt to encode the full constraining power of the BAO, as we are not taking into account current and past galaxy surveys' results, we do not consider overlapping redshift bins among surveys and we are discarding the distance information encoded in the quadrupole. 
Therefore the $> 3\sigma$ detection of acceleration we found is not the strongest one could obtain with the LP.
This should not be compared to the results of analyses performed with standard BAO techniques \cite{2013MNRAS.436.1674A, 2015PhRvD..92l3516A, 2020PhRvL.124v1301N, 2021PhRvD.103h3533A},
because those are obtained keeping cosmological and other phenomenological parameters fixed to fiducial values in flat-$\Lambda$CDM (i.e.~they are not derived using PG-BAO distance measures). 

The cosmological forecasts presented in this manuscript underline a serious overlooked issue of standard BAO analyses and suggest the LP standard ruler as a promising candidate to address it.
BAO constraints on cosmological parameters allow wide ranges of parameter values.
However, traditional BAO methodologies, in addition to not satisfying the Purely-Geometric-BAO necessary conditions, rest on hazardous extrapolations in parameter regions far from the fiducial flat-$\Lambda$CDM values they employ.
For the LP-inferred distances, a preliminary investigation successfully addressed this issue \cite{2020PhRvD.101h3517O}. 
That analysis needs to be extended by running N-body simulations in a wide enough parameter range; this will be the subject of future work.

If galaxy surveys are going to serve for truly independent measurements of cosmological observables, and independent inference of cosmological parameters, then 
galaxy survey data analysis pipelines --- in going from raw data to the final galaxy catalogs, from catalogs to distances, and from distances to parameters --- need to be extremely cautious in all the intervening fiducial cosmology assumptions. 
All of them need to be scrutinized if measured BAO distances are to be used, not just to improve parameter fits in standard $\Lambda$CDM, but to test the underlying cosmological model far from the fiducial parameter values and to explore modifications of the standard cosmological model. We plan to investigate this subject further.

\begin{acknowledgments} 
We thank Pier-Stefano Corasaniti for having suggested the present analysis and for many discussions. We thank Benjamin Wandelt for useful discussions on the Fisher information matrix. 
We thank Ravi Sheth and Idit Zehavi for discussions. 
SA thank Chris Blake for having underlined the 4MOST galaxy survey is relevant for BAO distances measures.
SA and AR were supported in part by the project ``Combining Cosmic Microwave Background and Large Scale Structure data: An Integrated Approach for Addressing Fundamental Questions in Cosmology,'' funded by the MIUR Progetti di Rilevante Interesse Nazionale (PRIN) Bando 2017, Grant No. 2017YJYZAH.
AR acknowledges funding from Italian Ministry of Education, University and Research (MIUR) through the ``Dipartimenti di eccellenza'' project Science of the Universe. 
GDS is partly supported by US Department of Energy grant DE-SC0009946 to the particle astrophysics theory group at CWRU.
\end{acknowledgments}

\phantom{.}\newline
\appendix
\section{DETAILS OF THE LINEAR POINT ESTIMATION FROM CF-MOCKS}
\label{appendix:LPestimation}

In this appendix we report technical details and choices that regard the Linear Point estimation procedure explained in Section \ref{DetProb}. We recall that, for each redshift survey and bin, we have 1000 synthetic CF realizations.

First, to verify that the $\chi_{\rm min}^2$ distribution (obtained from the polynomial fitting to the 1000 synthetic CF realizations) is consistent with the expected $\chi^2$ distribution, we perform the Kolmogorov-Smirnov test which returns a $p$-value. We require $p \geq 0.01$. We then ask the mean of the estimated LP to be within  $0.2 \times \sigma_{LP}$ w.r.t.~the true LP value computed from Eq.~(\ref{nl:xi}).

For each redshift survey and bin we furthermore check the Gaussianity of the LP best-fit distribution. In this case the Kolmogorov-Smirnov test returns $p$-values as low as 0.0001 for few redshift bins. This is not a problem per se, however in Section \ref{sec:Fisher} we assume a Gaussian likelihood to compare LP synthetic data to cosmological theoretical predictions, thus a high level of non-Gaussianity could invalidate this assumption. In this respect, first notice that we only employ the selected surveys and redshift bins reported in Table \ref{tab:selected}. Among those only the $\bar{z}=2.55$ redshift bin returns a p-value smaller than $0.01$. Therefore we did further tests to characterize the degree of non-Gaussianity and we found the following. The left and right errors are $30\%$ different, compared to the up to $20\%$ difference for $p \geq 0.01$ (i.e.~Gaussian) cases. We also split the $\sim 970$ mocks in three subsamples finding that all of them are compatible with a Gaussian distribution. We therefore conclude the degree of non-Gaussianity is probably not severe enough to invalidate our Gaussian likelihood assumption. A more careful investigation of this issue is beyond the scope of this manuscript.

Finally, in \cite{2018PhRvD..98b3527A} it is explained that if the LP-detection probability is smaller than $100\%$, the CF covariance needs to be re-calculated from the mocks where the LP is detected. However, in \cite{2018PhRvD..98b3527A} we found that the re-calculation of the covariance did not impact the results. Therefore in this manuscript we do not re-calculate the CF covariance assuming the same behavior applies to our cases. Nevertheless, for real observed data analysis the more rigorous re-computation of the CF covariance needs to be implemented.

\bibliography{MyBib}

\begin{thebibliography}{61}
\expandafter\ifx\csname natexlab\endcsname\relax\def\natexlab#1{#1}\fi
\expandafter\ifx\csname bibnamefont\endcsname\relax
  \def\bibnamefont#1{#1}\fi
\expandafter\ifx\csname bibfnamefont\endcsname\relax
  \def\bibfnamefont#1{#1}\fi
\expandafter\ifx\csname citenamefont\endcsname\relax
  \def\citenamefont#1{#1}\fi
\expandafter\ifx\csname url\endcsname\relax
  \def\url#1{\texttt{#1}}\fi
\expandafter\ifx\csname urlprefix\endcsname\relax\def\urlprefix{URL }\fi
\providecommand{\bibinfo}[2]{#2}
\providecommand{\eprint}[2][]{\url{#2}}

\bibitem[{\citenamefont{Perlmutter et~al.}(1999)}]{Perlmutter:1998np}
\bibinfo{author}{\bibfnamefont{S.}~\bibnamefont{Perlmutter}}
  \bibnamefont{et~al.} (\bibinfo{collaboration}{Supernova Cosmology Project}),
  \bibinfo{journal}{Astrophys.J.} \textbf{\bibinfo{volume}{517}},
  \bibinfo{pages}{565} (\bibinfo{year}{1999}), \eprint{astro-ph/9812133}.

\bibitem[{\citenamefont{Riess et~al.}(1998)}]{Riess:1998cb}
\bibinfo{author}{\bibfnamefont{A.~G.} \bibnamefont{Riess}} \bibnamefont{et~al.}
  (\bibinfo{collaboration}{Supernova Search Team}),
  \bibinfo{journal}{Astron.J.} \textbf{\bibinfo{volume}{116}},
  \bibinfo{pages}{1009} (\bibinfo{year}{1998}), \eprint{astro-ph/9805201}.

\bibitem[{\citenamefont{Bassett and Hlozek}(2009)}]{Bassett:2009mm}
\bibinfo{author}{\bibfnamefont{B.~A.} \bibnamefont{Bassett}} \bibnamefont{and}
  \bibinfo{author}{\bibfnamefont{R.}~\bibnamefont{Hlozek}}
  (\bibinfo{year}{2009}), \eprint{0910.5224}.

\bibitem[{\citenamefont{{Shanks} et~al.}(1987)\citenamefont{{Shanks}, {Fong},
  {Boyle}, and {Peterson}}}]{1987MNRAS.227..739S}
\bibinfo{author}{\bibfnamefont{T.}~\bibnamefont{{Shanks}}},
  \bibinfo{author}{\bibfnamefont{R.}~\bibnamefont{{Fong}}},
  \bibinfo{author}{\bibfnamefont{B.~J.} \bibnamefont{{Boyle}}},
  \bibnamefont{and} \bibinfo{author}{\bibfnamefont{B.~A.}
  \bibnamefont{{Peterson}}}, \bibinfo{journal}{\mnras}
  \textbf{\bibinfo{volume}{227}}, \bibinfo{pages}{739} (\bibinfo{year}{1987}).

\bibitem[{\citenamefont{{Smith} et~al.}(2008)\citenamefont{{Smith},
  {Scoccimarro}, and {Sheth}}}]{2008PhRvD..77d3525S}
\bibinfo{author}{\bibfnamefont{R.~E.} \bibnamefont{{Smith}}},
  \bibinfo{author}{\bibfnamefont{R.}~\bibnamefont{{Scoccimarro}}},
  \bibnamefont{and} \bibinfo{author}{\bibfnamefont{R.~K.}
  \bibnamefont{{Sheth}}}, \bibinfo{journal}{Phys. Rev. D}
  \textbf{\bibinfo{volume}{77}}, \bibinfo{eid}{043525} (\bibinfo{year}{2008}),
  \eprint{astro-ph/0703620}.

\bibitem[{\citenamefont{Spergel et~al.}(2007)}]{Spergel:2006hy}
\bibinfo{author}{\bibfnamefont{D.~N.} \bibnamefont{Spergel}}
  \bibnamefont{et~al.} (\bibinfo{collaboration}{WMAP}),
  \bibinfo{journal}{Astrophys. J. Suppl.} \textbf{\bibinfo{volume}{170}},
  \bibinfo{pages}{377} (\bibinfo{year}{2007}), \eprint{astro-ph/0603449}.

\bibitem[{\citenamefont{{Planck Collaboration}
  et~al.}(2020)\citenamefont{{Planck Collaboration}, {Aghanim}, {Akrami},
  {Ashdown}, {Aumont}, {Baccigalupi}, {Ballardini}, {Banday}, {Barreiro},
  {Bartolo} et~al.}}]{2020A&A...641A...6P}
\bibinfo{author}{\bibnamefont{{Planck Collaboration}}},
  \bibinfo{author}{\bibfnamefont{N.}~\bibnamefont{{Aghanim}}},
  \bibinfo{author}{\bibfnamefont{Y.}~\bibnamefont{{Akrami}}},
  \bibinfo{author}{\bibfnamefont{M.}~\bibnamefont{{Ashdown}}},
  \bibinfo{author}{\bibfnamefont{J.}~\bibnamefont{{Aumont}}},
  \bibinfo{author}{\bibfnamefont{C.}~\bibnamefont{{Baccigalupi}}},
  \bibinfo{author}{\bibfnamefont{M.}~\bibnamefont{{Ballardini}}},
  \bibinfo{author}{\bibfnamefont{A.~J.} \bibnamefont{{Banday}}},
  \bibinfo{author}{\bibfnamefont{R.~B.} \bibnamefont{{Barreiro}}},
  \bibinfo{author}{\bibfnamefont{N.}~\bibnamefont{{Bartolo}}},
  \bibnamefont{et~al.}, \bibinfo{journal}{\aap} \textbf{\bibinfo{volume}{641}},
  \bibinfo{eid}{A6} (\bibinfo{year}{2020}), \eprint{1807.06209}.

\bibitem[{\citenamefont{{Seo} et~al.}(2008)\citenamefont{{Seo}, {Siegel},
  {Eisenstein}, and {White}}}]{2008ApJ...686...13S}
\bibinfo{author}{\bibfnamefont{H.-J.} \bibnamefont{{Seo}}},
  \bibinfo{author}{\bibfnamefont{E.~R.} \bibnamefont{{Siegel}}},
  \bibinfo{author}{\bibfnamefont{D.~J.} \bibnamefont{{Eisenstein}}},
  \bibnamefont{and} \bibinfo{author}{\bibfnamefont{M.}~\bibnamefont{{White}}},
  \bibinfo{journal}{\apj} \textbf{\bibinfo{volume}{686}}, \bibinfo{pages}{13}
  (\bibinfo{year}{2008}), \eprint{0805.0117}.

\bibitem[{\citenamefont{{Xu} et~al.}(2012)\citenamefont{{Xu}, {Padmanabhan},
  {Eisenstein}, {Mehta}, and {Cuesta}}}]{2012MNRAS.427.2146X}
\bibinfo{author}{\bibfnamefont{X.}~\bibnamefont{{Xu}}},
  \bibinfo{author}{\bibfnamefont{N.}~\bibnamefont{{Padmanabhan}}},
  \bibinfo{author}{\bibfnamefont{D.~J.} \bibnamefont{{Eisenstein}}},
  \bibinfo{author}{\bibfnamefont{K.~T.} \bibnamefont{{Mehta}}},
  \bibnamefont{and} \bibinfo{author}{\bibfnamefont{A.~J.}
  \bibnamefont{{Cuesta}}}, \bibinfo{journal}{\mnras}
  \textbf{\bibinfo{volume}{427}}, \bibinfo{pages}{2146} (\bibinfo{year}{2012}),
  \eprint{1202.0091}.

\bibitem[{\citenamefont{{Carter} et~al.}(2020)\citenamefont{{Carter},
  {Beutler}, {Percival}, {DeRose}, {Wechsler}, and
  {Zhao}}}]{2020MNRAS.494.2076C}
\bibinfo{author}{\bibfnamefont{P.}~\bibnamefont{{Carter}}},
  \bibinfo{author}{\bibfnamefont{F.}~\bibnamefont{{Beutler}}},
  \bibinfo{author}{\bibfnamefont{W.~J.} \bibnamefont{{Percival}}},
  \bibinfo{author}{\bibfnamefont{J.}~\bibnamefont{{DeRose}}},
  \bibinfo{author}{\bibfnamefont{R.~H.} \bibnamefont{{Wechsler}}},
  \bibnamefont{and} \bibinfo{author}{\bibfnamefont{C.}~\bibnamefont{{Zhao}}},
  \bibinfo{journal}{\mnras} \textbf{\bibinfo{volume}{494}},
  \bibinfo{pages}{2076} (\bibinfo{year}{2020}), \eprint{1906.03035}.

\bibitem[{\citenamefont{{Anselmi} et~al.}(2019)\citenamefont{{Anselmi},
  {Corasaniti}, {Sanchez}, {Starkman}, {Sheth}, and
  {Zehavi}}}]{2019PhRvD..99l3515A}
\bibinfo{author}{\bibfnamefont{S.}~\bibnamefont{{Anselmi}}},
  \bibinfo{author}{\bibfnamefont{P.-S.} \bibnamefont{{Corasaniti}}},
  \bibinfo{author}{\bibfnamefont{A.~G.} \bibnamefont{{Sanchez}}},
  \bibinfo{author}{\bibfnamefont{G.~D.} \bibnamefont{{Starkman}}},
  \bibinfo{author}{\bibfnamefont{R.~K.} \bibnamefont{{Sheth}}},
  \bibnamefont{and} \bibinfo{author}{\bibfnamefont{I.}~\bibnamefont{{Zehavi}}},
  \bibinfo{journal}{\prd} \textbf{\bibinfo{volume}{99}}, \bibinfo{eid}{123515}
  (\bibinfo{year}{2019}), \eprint{1811.12312}.

\bibitem[{\citenamefont{{Bernal} et~al.}(2020)\citenamefont{{Bernal}, {Smith},
  {Boddy}, and {Kamionkowski}}}]{2020PhRvD.102l3515B}
\bibinfo{author}{\bibfnamefont{J.~L.} \bibnamefont{{Bernal}}},
  \bibinfo{author}{\bibfnamefont{T.~L.} \bibnamefont{{Smith}}},
  \bibinfo{author}{\bibfnamefont{K.~K.} \bibnamefont{{Boddy}}},
  \bibnamefont{and}
  \bibinfo{author}{\bibfnamefont{M.}~\bibnamefont{{Kamionkowski}}},
  \bibinfo{journal}{\prd} \textbf{\bibinfo{volume}{102}}, \bibinfo{eid}{123515}
  (\bibinfo{year}{2020}), \eprint{2004.07263}.

\bibitem[{\citenamefont{{Beutler} et~al.}(2017)\citenamefont{{Beutler}, {Seo},
  {Saito}, {Chuang}, {Cuesta}, {Eisenstein}, {Gil-Mar{\'\i}n}, {Grieb}, {Hand},
  {Kitaura} et~al.}}]{2017MNRAS.466.2242B}
\bibinfo{author}{\bibfnamefont{F.}~\bibnamefont{{Beutler}}},
  \bibinfo{author}{\bibfnamefont{H.-J.} \bibnamefont{{Seo}}},
  \bibinfo{author}{\bibfnamefont{S.}~\bibnamefont{{Saito}}},
  \bibinfo{author}{\bibfnamefont{C.-H.} \bibnamefont{{Chuang}}},
  \bibinfo{author}{\bibfnamefont{A.~J.} \bibnamefont{{Cuesta}}},
  \bibinfo{author}{\bibfnamefont{D.~J.} \bibnamefont{{Eisenstein}}},
  \bibinfo{author}{\bibfnamefont{H.}~\bibnamefont{{Gil-Mar{\'\i}n}}},
  \bibinfo{author}{\bibfnamefont{J.~N.} \bibnamefont{{Grieb}}},
  \bibinfo{author}{\bibfnamefont{N.}~\bibnamefont{{Hand}}},
  \bibinfo{author}{\bibfnamefont{F.-S.} \bibnamefont{{Kitaura}}},
  \bibnamefont{et~al.}, \bibinfo{journal}{\mnras}
  \textbf{\bibinfo{volume}{466}}, \bibinfo{pages}{2242} (\bibinfo{year}{2017}),
  \eprint{1607.03150}.

\bibitem[{\citenamefont{{Brieden} et~al.}(2021)\citenamefont{{Brieden},
  {Gil-Mar{\'\i}n}, and {Verde}}}]{2021JCAP...12..054B}
\bibinfo{author}{\bibfnamefont{S.}~\bibnamefont{{Brieden}}},
  \bibinfo{author}{\bibfnamefont{H.}~\bibnamefont{{Gil-Mar{\'\i}n}}},
  \bibnamefont{and} \bibinfo{author}{\bibfnamefont{L.}~\bibnamefont{{Verde}}},
  \bibinfo{journal}{\jcap} \textbf{\bibinfo{volume}{2021}}, \bibinfo{eid}{054}
  (\bibinfo{year}{2021}), \eprint{2106.07641}.

\bibitem[{\citenamefont{{Eisenstein} et~al.}(2007)\citenamefont{{Eisenstein},
  {Seo}, {Sirko}, and {Spergel}}}]{2007ApJ...664..675E}
\bibinfo{author}{\bibfnamefont{D.~J.} \bibnamefont{{Eisenstein}}},
  \bibinfo{author}{\bibfnamefont{H.-J.} \bibnamefont{{Seo}}},
  \bibinfo{author}{\bibfnamefont{E.}~\bibnamefont{{Sirko}}}, \bibnamefont{and}
  \bibinfo{author}{\bibfnamefont{D.~N.} \bibnamefont{{Spergel}}},
  \bibinfo{journal}{\apj} \textbf{\bibinfo{volume}{664}}, \bibinfo{pages}{675}
  (\bibinfo{year}{2007}), \eprint{astro-ph/0604362}.

\bibitem[{\citenamefont{{Padmanabhan} et~al.}(2012)\citenamefont{{Padmanabhan},
  {Xu}, {Eisenstein}, {Scalzo}, {Cuesta}, {Mehta}, and
  {Kazin}}}]{2012MNRAS.427.2132P}
\bibinfo{author}{\bibfnamefont{N.}~\bibnamefont{{Padmanabhan}}},
  \bibinfo{author}{\bibfnamefont{X.}~\bibnamefont{{Xu}}},
  \bibinfo{author}{\bibfnamefont{D.~J.} \bibnamefont{{Eisenstein}}},
  \bibinfo{author}{\bibfnamefont{R.}~\bibnamefont{{Scalzo}}},
  \bibinfo{author}{\bibfnamefont{A.~J.} \bibnamefont{{Cuesta}}},
  \bibinfo{author}{\bibfnamefont{K.~T.} \bibnamefont{{Mehta}}},
  \bibnamefont{and} \bibinfo{author}{\bibfnamefont{E.}~\bibnamefont{{Kazin}}},
  \bibinfo{journal}{\mnras} \textbf{\bibinfo{volume}{427}},
  \bibinfo{pages}{2132} (\bibinfo{year}{2012}), \eprint{1202.0090}.

\bibitem[{\citenamefont{{Nikakhtar}
  et~al.}(2021{\natexlab{a}})\citenamefont{{Nikakhtar}, {Sheth}, and
  {Zehavi}}}]{2021PhRvD.104d3530N}
\bibinfo{author}{\bibfnamefont{F.}~\bibnamefont{{Nikakhtar}}},
  \bibinfo{author}{\bibfnamefont{R.~K.} \bibnamefont{{Sheth}}},
  \bibnamefont{and} \bibinfo{author}{\bibfnamefont{I.}~\bibnamefont{{Zehavi}}},
  \bibinfo{journal}{\prd} \textbf{\bibinfo{volume}{104}}, \bibinfo{eid}{043530}
  (\bibinfo{year}{2021}{\natexlab{a}}), \eprint{2101.08376}.

\bibitem[{\citenamefont{{Levy} et~al.}(2021)\citenamefont{{Levy}, {Mohayaee},
  and {von Hausegger}}}]{2021MNRAS.506.1165L}
\bibinfo{author}{\bibfnamefont{B.}~\bibnamefont{{Levy}}},
  \bibinfo{author}{\bibfnamefont{R.}~\bibnamefont{{Mohayaee}}},
  \bibnamefont{and} \bibinfo{author}{\bibfnamefont{S.}~\bibnamefont{{von
  Hausegger}}}, \bibinfo{journal}{\mnras} \textbf{\bibinfo{volume}{506}},
  \bibinfo{pages}{1165} (\bibinfo{year}{2021}), \eprint{2012.09074}.

\bibitem[{\citenamefont{{von Hausegger} et~al.}(2022)\citenamefont{{von
  Hausegger}, {L{\'e}vy}, and {Mohayaee}}}]{2022PhRvL.128t1302V}
\bibinfo{author}{\bibfnamefont{S.}~\bibnamefont{{von Hausegger}}},
  \bibinfo{author}{\bibfnamefont{B.}~\bibnamefont{{L{\'e}vy}}},
  \bibnamefont{and}
  \bibinfo{author}{\bibfnamefont{R.}~\bibnamefont{{Mohayaee}}},
  \bibinfo{journal}{\prl} \textbf{\bibinfo{volume}{128}}, \bibinfo{eid}{201302}
  (\bibinfo{year}{2022}), \eprint{2110.08868}.

\bibitem[{\citenamefont{{Nikakhtar} et~al.}(2022)\citenamefont{{Nikakhtar},
  {Sheth}, {L{\'e}vy}, and {Mohayaee}}}]{2022arXiv220301868N}
\bibinfo{author}{\bibfnamefont{F.}~\bibnamefont{{Nikakhtar}}},
  \bibinfo{author}{\bibfnamefont{R.~K.} \bibnamefont{{Sheth}}},
  \bibinfo{author}{\bibfnamefont{B.}~\bibnamefont{{L{\'e}vy}}},
  \bibnamefont{and}
  \bibinfo{author}{\bibfnamefont{R.}~\bibnamefont{{Mohayaee}}},
  \bibinfo{journal}{arXiv e-prints} \bibinfo{eid}{arXiv:2203.01868}
  (\bibinfo{year}{2022}), \eprint{2203.01868}.

\bibitem[{\citenamefont{{Osato} et~al.}(2019)\citenamefont{{Osato},
  {Nishimichi}, {Bernardeau}, and {Taruya}}}]{2019PhRvD..99f3530O}
\bibinfo{author}{\bibfnamefont{K.}~\bibnamefont{{Osato}}},
  \bibinfo{author}{\bibfnamefont{T.}~\bibnamefont{{Nishimichi}}},
  \bibinfo{author}{\bibfnamefont{F.}~\bibnamefont{{Bernardeau}}},
  \bibnamefont{and} \bibinfo{author}{\bibfnamefont{A.}~\bibnamefont{{Taruya}}},
  \bibinfo{journal}{\prd} \textbf{\bibinfo{volume}{99}}, \bibinfo{eid}{063530}
  (\bibinfo{year}{2019}), \eprint{1810.10104}.

\bibitem[{\citenamefont{{Oddo} et~al.}(2021)\citenamefont{{Oddo}, {Rizzo},
  {Sefusatti}, {Porciani}, and {Monaco}}}]{2021JCAP...11..038O}
\bibinfo{author}{\bibfnamefont{A.}~\bibnamefont{{Oddo}}},
  \bibinfo{author}{\bibfnamefont{F.}~\bibnamefont{{Rizzo}}},
  \bibinfo{author}{\bibfnamefont{E.}~\bibnamefont{{Sefusatti}}},
  \bibinfo{author}{\bibfnamefont{C.}~\bibnamefont{{Porciani}}},
  \bibnamefont{and} \bibinfo{author}{\bibfnamefont{P.}~\bibnamefont{{Monaco}}},
  \bibinfo{journal}{\jcap} \textbf{\bibinfo{volume}{2021}}, \bibinfo{eid}{038}
  (\bibinfo{year}{2021}), \eprint{2108.03204}.

\bibitem[{\citenamefont{{Alkhanishvili}
  et~al.}(2022)\citenamefont{{Alkhanishvili}, {Porciani}, {Sefusatti},
  {Biagetti}, {Lazanu}, {Oddo}, and {Yankelevich}}}]{2022MNRAS.tmp..572A}
\bibinfo{author}{\bibfnamefont{D.}~\bibnamefont{{Alkhanishvili}}},
  \bibinfo{author}{\bibfnamefont{C.}~\bibnamefont{{Porciani}}},
  \bibinfo{author}{\bibfnamefont{E.}~\bibnamefont{{Sefusatti}}},
  \bibinfo{author}{\bibfnamefont{M.}~\bibnamefont{{Biagetti}}},
  \bibinfo{author}{\bibfnamefont{A.}~\bibnamefont{{Lazanu}}},
  \bibinfo{author}{\bibfnamefont{A.}~\bibnamefont{{Oddo}}}, \bibnamefont{and}
  \bibinfo{author}{\bibfnamefont{V.}~\bibnamefont{{Yankelevich}}},
  \bibinfo{journal}{\mnras}  (\bibinfo{year}{2022}), \eprint{2107.08054}.

\bibitem[{\citenamefont{{Anselmi} et~al.}(2016)\citenamefont{{Anselmi},
  {Starkman}, and {Sheth}}}]{2016MNRAS.455.2474A}
\bibinfo{author}{\bibfnamefont{S.}~\bibnamefont{{Anselmi}}},
  \bibinfo{author}{\bibfnamefont{G.~D.} \bibnamefont{{Starkman}}},
  \bibnamefont{and} \bibinfo{author}{\bibfnamefont{R.~K.}
  \bibnamefont{{Sheth}}}, \bibinfo{journal}{\mnras}
  \textbf{\bibinfo{volume}{455}}, \bibinfo{pages}{2474} (\bibinfo{year}{2016}),
  \eprint{1508.01170}.

\bibitem[{\citenamefont{{Anselmi}
  et~al.}(2018{\natexlab{a}})\citenamefont{{Anselmi}, {Starkman}, {Corasaniti},
  {Sheth}, and {Zehavi}}}]{2018PhRvL.121b1302A}
\bibinfo{author}{\bibfnamefont{S.}~\bibnamefont{{Anselmi}}},
  \bibinfo{author}{\bibfnamefont{G.~D.} \bibnamefont{{Starkman}}},
  \bibinfo{author}{\bibfnamefont{P.-S.} \bibnamefont{{Corasaniti}}},
  \bibinfo{author}{\bibfnamefont{R.~K.} \bibnamefont{{Sheth}}},
  \bibnamefont{and} \bibinfo{author}{\bibfnamefont{I.}~\bibnamefont{{Zehavi}}},
  \bibinfo{journal}{\prl} \textbf{\bibinfo{volume}{121}}, \bibinfo{eid}{021302}
  (\bibinfo{year}{2018}{\natexlab{a}}).

\bibitem[{\citenamefont{{Parimbelli} et~al.}(2021)\citenamefont{{Parimbelli},
  {Anselmi}, {Viel}, {Carbone}, {Villaescusa-Navarro}, {Corasaniti}, {Rasera},
  {Sheth}, {Starkman}, and {Zehavi}}}]{2021JCAP...01..009P}
\bibinfo{author}{\bibfnamefont{G.}~\bibnamefont{{Parimbelli}}},
  \bibinfo{author}{\bibfnamefont{S.}~\bibnamefont{{Anselmi}}},
  \bibinfo{author}{\bibfnamefont{M.}~\bibnamefont{{Viel}}},
  \bibinfo{author}{\bibfnamefont{C.}~\bibnamefont{{Carbone}}},
  \bibinfo{author}{\bibfnamefont{F.}~\bibnamefont{{Villaescusa-Navarro}}},
  \bibinfo{author}{\bibfnamefont{P.~S.} \bibnamefont{{Corasaniti}}},
  \bibinfo{author}{\bibfnamefont{Y.}~\bibnamefont{{Rasera}}},
  \bibinfo{author}{\bibfnamefont{R.}~\bibnamefont{{Sheth}}},
  \bibinfo{author}{\bibfnamefont{G.~D.} \bibnamefont{{Starkman}}},
  \bibnamefont{and} \bibinfo{author}{\bibfnamefont{I.}~\bibnamefont{{Zehavi}}},
  \bibinfo{journal}{\jcap} \textbf{\bibinfo{volume}{2021}}, \bibinfo{eid}{009}
  (\bibinfo{year}{2021}), \eprint{2007.10345}.

\bibitem[{\citenamefont{{Camarena} and {Marra}}(2020)}]{2020MNRAS.495.2630C}
\bibinfo{author}{\bibfnamefont{D.}~\bibnamefont{{Camarena}}} \bibnamefont{and}
  \bibinfo{author}{\bibfnamefont{V.}~\bibnamefont{{Marra}}},
  \bibinfo{journal}{\mnras} \textbf{\bibinfo{volume}{495}},
  \bibinfo{pages}{2630} (\bibinfo{year}{2020}), \eprint{1910.14125}.

\bibitem[{\citenamefont{{Haridasu} et~al.}(2018)\citenamefont{{Haridasu},
  {Lukovi{\'c}}, and {Vittorio}}}]{2018JCAP...05..033H}
\bibinfo{author}{\bibfnamefont{B.~S.} \bibnamefont{{Haridasu}}},
  \bibinfo{author}{\bibfnamefont{V.~V.} \bibnamefont{{Lukovi{\'c}}}},
  \bibnamefont{and}
  \bibinfo{author}{\bibfnamefont{N.}~\bibnamefont{{Vittorio}}},
  \bibinfo{journal}{\jcap} \textbf{\bibinfo{volume}{5}}, \bibinfo{eid}{033}
  (\bibinfo{year}{2018}), \eprint{1711.03929}.

\bibitem[{\citenamefont{{O'Dwyer} et~al.}(2020)\citenamefont{{O'Dwyer},
  {Anselmi}, {Starkman}, {Corasaniti}, {Sheth}, and
  {Zehavi}}}]{2020PhRvD.101h3517O}
\bibinfo{author}{\bibfnamefont{M.}~\bibnamefont{{O'Dwyer}}},
  \bibinfo{author}{\bibfnamefont{S.}~\bibnamefont{{Anselmi}}},
  \bibinfo{author}{\bibfnamefont{G.~D.} \bibnamefont{{Starkman}}},
  \bibinfo{author}{\bibfnamefont{P.-S.} \bibnamefont{{Corasaniti}}},
  \bibinfo{author}{\bibfnamefont{R.~K.} \bibnamefont{{Sheth}}},
  \bibnamefont{and} \bibinfo{author}{\bibfnamefont{I.}~\bibnamefont{{Zehavi}}},
  \bibinfo{journal}{\prd} \textbf{\bibinfo{volume}{101}}, \bibinfo{eid}{083517}
  (\bibinfo{year}{2020}), \eprint{1910.10698}.

\bibitem[{\citenamefont{{Nielsen} et~al.}(2016)\citenamefont{{Nielsen},
  {Guffanti}, and {Sarkar}}}]{2016NatSR...635596N}
\bibinfo{author}{\bibfnamefont{J.~T.} \bibnamefont{{Nielsen}}},
  \bibinfo{author}{\bibfnamefont{A.}~\bibnamefont{{Guffanti}}},
  \bibnamefont{and} \bibinfo{author}{\bibfnamefont{S.}~\bibnamefont{{Sarkar}}},
  \bibinfo{journal}{Scientific Reports} \textbf{\bibinfo{volume}{6}},
  \bibinfo{eid}{35596} (\bibinfo{year}{2016}).

\bibitem[{\citenamefont{{Rahman} et~al.}(2021)\citenamefont{{Rahman}, {Trotta},
  {Boruah}, {Hudson}, and {van Dyk}}}]{2021arXiv210812497R}
\bibinfo{author}{\bibfnamefont{W.}~\bibnamefont{{Rahman}}},
  \bibinfo{author}{\bibfnamefont{R.}~\bibnamefont{{Trotta}}},
  \bibinfo{author}{\bibfnamefont{S.~S.} \bibnamefont{{Boruah}}},
  \bibinfo{author}{\bibfnamefont{M.~J.} \bibnamefont{{Hudson}}},
  \bibnamefont{and} \bibinfo{author}{\bibfnamefont{D.~A.} \bibnamefont{{van
  Dyk}}}, \bibinfo{journal}{arXiv e-prints} \bibinfo{eid}{arXiv:2108.12497}
  (\bibinfo{year}{2021}), \eprint{2108.12497}.

\bibitem[{\citenamefont{{Glanville} et~al.}(2022)\citenamefont{{Glanville},
  {Howlett}, and {Davis}}}]{2022arXiv220505892G}
\bibinfo{author}{\bibfnamefont{A.}~\bibnamefont{{Glanville}}},
  \bibinfo{author}{\bibfnamefont{C.}~\bibnamefont{{Howlett}}},
  \bibnamefont{and} \bibinfo{author}{\bibfnamefont{T.~M.}
  \bibnamefont{{Davis}}}, \bibinfo{journal}{arXiv e-prints}
  \bibinfo{eid}{arXiv:2205.05892} (\bibinfo{year}{2022}), \eprint{2205.05892}.

\bibitem[{\citenamefont{{Heinesen} et~al.}(2020)\citenamefont{{Heinesen},
  {Blake}, and {Wiltshire}}}]{2020JCAP...01..038H}
\bibinfo{author}{\bibfnamefont{A.}~\bibnamefont{{Heinesen}}},
  \bibinfo{author}{\bibfnamefont{C.}~\bibnamefont{{Blake}}}, \bibnamefont{and}
  \bibinfo{author}{\bibfnamefont{D.~L.} \bibnamefont{{Wiltshire}}},
  \bibinfo{journal}{\jcap} \textbf{\bibinfo{volume}{2020}}, \bibinfo{eid}{038}
  (\bibinfo{year}{2020}), \eprint{1908.11508}.

\bibitem[{\citenamefont{Kaiser}(1987)}]{Kaiser:1987qv}
\bibinfo{author}{\bibfnamefont{N.}~\bibnamefont{Kaiser}},
  \bibinfo{journal}{Mon.Not.Roy.Astron.Soc.} \textbf{\bibinfo{volume}{227}},
  \bibinfo{pages}{1} (\bibinfo{year}{1987}).

\bibitem[{\citenamefont{{Smith}}(2009)}]{2009MNRAS.400..851S}
\bibinfo{author}{\bibfnamefont{R.~E.} \bibnamefont{{Smith}}},
  \bibinfo{journal}{\mnras} \textbf{\bibinfo{volume}{400}},
  \bibinfo{pages}{851} (\bibinfo{year}{2009}), \eprint{0810.1960}.

\bibitem[{\citenamefont{{Grieb} et~al.}(2016)\citenamefont{{Grieb},
  {S{\'a}nchez}, {Salazar-Albornoz}, and {Dalla
  Vecchia}}}]{2016MNRAS.457.1577G}
\bibinfo{author}{\bibfnamefont{J.~N.} \bibnamefont{{Grieb}}},
  \bibinfo{author}{\bibfnamefont{A.~G.} \bibnamefont{{S{\'a}nchez}}},
  \bibinfo{author}{\bibfnamefont{S.}~\bibnamefont{{Salazar-Albornoz}}},
  \bibnamefont{and} \bibinfo{author}{\bibfnamefont{C.}~\bibnamefont{{Dalla
  Vecchia}}}, \bibinfo{journal}{\mnras} \textbf{\bibinfo{volume}{457}},
  \bibinfo{pages}{1577} (\bibinfo{year}{2016}), \eprint{1509.04293}.

\bibitem[{\citenamefont{{Anselmi}
  et~al.}(2018{\natexlab{b}})\citenamefont{{Anselmi}, {Corasaniti}, {Starkman},
  {Sheth}, and {Zehavi}}}]{2018PhRvD..98b3527A}
\bibinfo{author}{\bibfnamefont{S.}~\bibnamefont{{Anselmi}}},
  \bibinfo{author}{\bibfnamefont{P.-S.} \bibnamefont{{Corasaniti}}},
  \bibinfo{author}{\bibfnamefont{G.~D.} \bibnamefont{{Starkman}}},
  \bibinfo{author}{\bibfnamefont{R.~K.} \bibnamefont{{Sheth}}},
  \bibnamefont{and} \bibinfo{author}{\bibfnamefont{I.}~\bibnamefont{{Zehavi}}},
  \bibinfo{journal}{\prd} \textbf{\bibinfo{volume}{98}}, \bibinfo{eid}{023527}
  (\bibinfo{year}{2018}{\natexlab{b}}), \eprint{1711.09063}.

\bibitem[{\citenamefont{{S{\'a}nchez} et~al.}(2013)\citenamefont{{S{\'a}nchez},
  {Alonso}, {S{\'a}nchez}, {Garc{\'{\i}}a-Bellido}, and
  {Sevilla}}}]{2013MNRAS.434.2008S}
\bibinfo{author}{\bibfnamefont{E.}~\bibnamefont{{S{\'a}nchez}}},
  \bibinfo{author}{\bibfnamefont{D.}~\bibnamefont{{Alonso}}},
  \bibinfo{author}{\bibfnamefont{F.~J.} \bibnamefont{{S{\'a}nchez}}},
  \bibinfo{author}{\bibfnamefont{J.}~\bibnamefont{{Garc{\'{\i}}a-Bellido}}},
  \bibnamefont{and}
  \bibinfo{author}{\bibfnamefont{I.}~\bibnamefont{{Sevilla}}},
  \bibinfo{journal}{\mnras} \textbf{\bibinfo{volume}{434}},
  \bibinfo{pages}{2008} (\bibinfo{year}{2013}), \eprint{1210.6446}.

\bibitem[{\citenamefont{{S{\'a}nchez} et~al.}(2011)\citenamefont{{S{\'a}nchez},
  {Carnero}, {Garc{\'\i}a-Bellido}, {Gazta{\~n}aga}, {de Simoni}, {Crocce},
  {Cabr{\'e}}, {Fosalba}, and {Alonso}}}]{2011MNRAS.411..277S}
\bibinfo{author}{\bibfnamefont{E.}~\bibnamefont{{S{\'a}nchez}}},
  \bibinfo{author}{\bibfnamefont{A.}~\bibnamefont{{Carnero}}},
  \bibinfo{author}{\bibfnamefont{J.}~\bibnamefont{{Garc{\'\i}a-Bellido}}},
  \bibinfo{author}{\bibfnamefont{E.}~\bibnamefont{{Gazta{\~n}aga}}},
  \bibinfo{author}{\bibfnamefont{F.}~\bibnamefont{{de Simoni}}},
  \bibinfo{author}{\bibfnamefont{M.}~\bibnamefont{{Crocce}}},
  \bibinfo{author}{\bibfnamefont{A.}~\bibnamefont{{Cabr{\'e}}}},
  \bibinfo{author}{\bibfnamefont{P.}~\bibnamefont{{Fosalba}}},
  \bibnamefont{and} \bibinfo{author}{\bibfnamefont{D.}~\bibnamefont{{Alonso}}},
  \bibinfo{journal}{\mnras} \textbf{\bibinfo{volume}{411}},
  \bibinfo{pages}{277} (\bibinfo{year}{2011}), \eprint{1006.3226}.

\bibitem[{\citenamefont{{Marra} and {Chirinos
  Isidro}}(2019)}]{2019MNRAS.487.3419M}
\bibinfo{author}{\bibfnamefont{V.}~\bibnamefont{{Marra}}} \bibnamefont{and}
  \bibinfo{author}{\bibfnamefont{E.~G.} \bibnamefont{{Chirinos Isidro}}},
  \bibinfo{journal}{\mnras} \textbf{\bibinfo{volume}{487}},
  \bibinfo{pages}{3419} (\bibinfo{year}{2019}), \eprint{1808.10695}.

\bibitem[{\citenamefont{{Xu} et~al.}(2013)\citenamefont{{Xu}, {Cuesta},
  {Padmanabhan}, {Eisenstein}, and {McBride}}}]{2013MNRAS.431.2834X}
\bibinfo{author}{\bibfnamefont{X.}~\bibnamefont{{Xu}}},
  \bibinfo{author}{\bibfnamefont{A.~J.} \bibnamefont{{Cuesta}}},
  \bibinfo{author}{\bibfnamefont{N.}~\bibnamefont{{Padmanabhan}}},
  \bibinfo{author}{\bibfnamefont{D.~J.} \bibnamefont{{Eisenstein}}},
  \bibnamefont{and} \bibinfo{author}{\bibfnamefont{C.~K.}
  \bibnamefont{{McBride}}}, \bibinfo{journal}{\mnras}
  \textbf{\bibinfo{volume}{431}}, \bibinfo{pages}{2834} (\bibinfo{year}{2013}),
  \eprint{1206.6732}.

\bibitem[{\citenamefont{{Cuesta} et~al.}(2016)\citenamefont{{Cuesta},
  {Vargas-Maga{\~n}a}, {Beutler}, {Bolton}, {Brownstein}, {Eisenstein},
  {Gil-Mar{\'{\i}}n}, {Ho}, {McBride}, {Maraston}
  et~al.}}]{2016MNRAS.457.1770C}
\bibinfo{author}{\bibfnamefont{A.~J.} \bibnamefont{{Cuesta}}},
  \bibinfo{author}{\bibfnamefont{M.}~\bibnamefont{{Vargas-Maga{\~n}a}}},
  \bibinfo{author}{\bibfnamefont{F.}~\bibnamefont{{Beutler}}},
  \bibinfo{author}{\bibfnamefont{A.~S.} \bibnamefont{{Bolton}}},
  \bibinfo{author}{\bibfnamefont{J.~R.} \bibnamefont{{Brownstein}}},
  \bibinfo{author}{\bibfnamefont{D.~J.} \bibnamefont{{Eisenstein}}},
  \bibinfo{author}{\bibfnamefont{H.}~\bibnamefont{{Gil-Mar{\'{\i}}n}}},
  \bibinfo{author}{\bibfnamefont{S.}~\bibnamefont{{Ho}}},
  \bibinfo{author}{\bibfnamefont{C.~K.} \bibnamefont{{McBride}}},
  \bibinfo{author}{\bibfnamefont{C.}~\bibnamefont{{Maraston}}},
  \bibnamefont{et~al.}, \bibinfo{journal}{\mnras}
  \textbf{\bibinfo{volume}{457}}, \bibinfo{pages}{1770} (\bibinfo{year}{2016}),
  \eprint{1509.06371}.

\bibitem[{\citenamefont{{Nikakhtar}
  et~al.}(2021{\natexlab{b}})\citenamefont{{Nikakhtar}, {Sheth}, and
  {Zehavi}}}]{2021PhRvD.104f3504N}
\bibinfo{author}{\bibfnamefont{F.}~\bibnamefont{{Nikakhtar}}},
  \bibinfo{author}{\bibfnamefont{R.~K.} \bibnamefont{{Sheth}}},
  \bibnamefont{and} \bibinfo{author}{\bibfnamefont{I.}~\bibnamefont{{Zehavi}}},
  \bibinfo{journal}{\prd} \textbf{\bibinfo{volume}{104}}, \bibinfo{eid}{063504}
  (\bibinfo{year}{2021}{\natexlab{b}}), \eprint{2107.12537}.

\bibitem[{\citenamefont{{Huterer} and {Starkman}}(2003)}]{2003PhRvL..90c1301H}
\bibinfo{author}{\bibfnamefont{D.}~\bibnamefont{{Huterer}}} \bibnamefont{and}
  \bibinfo{author}{\bibfnamefont{G.}~\bibnamefont{{Starkman}}},
  \bibinfo{journal}{\prl} \textbf{\bibinfo{volume}{90}}, \bibinfo{eid}{031301}
  (\bibinfo{year}{2003}), \eprint{astro-ph/0207517}.

\bibitem[{\citenamefont{{Tegmark} et~al.}(1997)\citenamefont{{Tegmark},
  {Taylor}, and {Heavens}}}]{1997ApJ...480...22T}
\bibinfo{author}{\bibfnamefont{M.}~\bibnamefont{{Tegmark}}},
  \bibinfo{author}{\bibfnamefont{A.~N.} \bibnamefont{{Taylor}}},
  \bibnamefont{and} \bibinfo{author}{\bibfnamefont{A.~F.}
  \bibnamefont{{Heavens}}}, \bibinfo{journal}{\apj}
  \textbf{\bibinfo{volume}{480}}, \bibinfo{pages}{22} (\bibinfo{year}{1997}),
  \eprint{astro-ph/9603021}.

\bibitem[{\citenamefont{{Mukherjee} et~al.}(2006)\citenamefont{{Mukherjee},
  {Parkinson}, {Corasaniti}, {Liddle}, and {Kunz}}}]{2006MNRAS.369.1725M}
\bibinfo{author}{\bibfnamefont{P.}~\bibnamefont{{Mukherjee}}},
  \bibinfo{author}{\bibfnamefont{D.}~\bibnamefont{{Parkinson}}},
  \bibinfo{author}{\bibfnamefont{P.~S.} \bibnamefont{{Corasaniti}}},
  \bibinfo{author}{\bibfnamefont{A.~R.} \bibnamefont{{Liddle}}},
  \bibnamefont{and} \bibinfo{author}{\bibfnamefont{M.}~\bibnamefont{{Kunz}}},
  \bibinfo{journal}{\mnras} \textbf{\bibinfo{volume}{369}},
  \bibinfo{pages}{1725} (\bibinfo{year}{2006}), \eprint{astro-ph/0512484}.

\bibitem[{\citenamefont{{Planck Collaboration}
  et~al.}(2016)\citenamefont{{Planck Collaboration}, {Ade}, {Aghanim},
  {Arnaud}, {Ashdown}, {Aumont}, {Baccigalupi}, {Banday}, {Barreiro},
  {Bartlett} et~al.}}]{2016A&A...594A..13P}
\bibinfo{author}{\bibnamefont{{Planck Collaboration}}},
  \bibinfo{author}{\bibfnamefont{P.~A.~R.} \bibnamefont{{Ade}}},
  \bibinfo{author}{\bibfnamefont{N.}~\bibnamefont{{Aghanim}}},
  \bibinfo{author}{\bibfnamefont{M.}~\bibnamefont{{Arnaud}}},
  \bibinfo{author}{\bibfnamefont{M.}~\bibnamefont{{Ashdown}}},
  \bibinfo{author}{\bibfnamefont{J.}~\bibnamefont{{Aumont}}},
  \bibinfo{author}{\bibfnamefont{C.}~\bibnamefont{{Baccigalupi}}},
  \bibinfo{author}{\bibfnamefont{A.~J.} \bibnamefont{{Banday}}},
  \bibinfo{author}{\bibfnamefont{R.~B.} \bibnamefont{{Barreiro}}},
  \bibinfo{author}{\bibfnamefont{J.~G.} \bibnamefont{{Bartlett}}},
  \bibnamefont{et~al.}, \bibinfo{journal}{\aap} \textbf{\bibinfo{volume}{594}},
  \bibinfo{eid}{A13} (\bibinfo{year}{2016}), \eprint{1502.01589}.

\bibitem[{\citenamefont{{Euclid Collaboration}
  et~al.}(2020)\citenamefont{{Euclid Collaboration}, {Blanchard}, {Camera},
  {Carbone}, {Cardone}, {Casas}, {Clesse}, {Ili{\'c}}, {Kilbinger}, {Kitching}
  et~al.}}]{2020A&A...642A.191E}
\bibinfo{author}{\bibnamefont{{Euclid Collaboration}}},
  \bibinfo{author}{\bibfnamefont{A.}~\bibnamefont{{Blanchard}}},
  \bibinfo{author}{\bibfnamefont{S.}~\bibnamefont{{Camera}}},
  \bibinfo{author}{\bibfnamefont{C.}~\bibnamefont{{Carbone}}},
  \bibinfo{author}{\bibfnamefont{V.~F.} \bibnamefont{{Cardone}}},
  \bibinfo{author}{\bibfnamefont{S.}~\bibnamefont{{Casas}}},
  \bibinfo{author}{\bibfnamefont{S.}~\bibnamefont{{Clesse}}},
  \bibinfo{author}{\bibfnamefont{S.}~\bibnamefont{{Ili{\'c}}}},
  \bibinfo{author}{\bibfnamefont{M.}~\bibnamefont{{Kilbinger}}},
  \bibinfo{author}{\bibfnamefont{T.}~\bibnamefont{{Kitching}}},
  \bibnamefont{et~al.}, \bibinfo{journal}{\aap} \textbf{\bibinfo{volume}{642}},
  \bibinfo{eid}{A191} (\bibinfo{year}{2020}), \eprint{1910.09273}.

\bibitem[{\citenamefont{{DESI Collaboration} et~al.}(2016)\citenamefont{{DESI
  Collaboration}, {Aghamousa}, {Aguilar}, {Ahlen}, {Alam}, {Allen}, {Allende
  Prieto}, {Annis}, {Bailey}, {Balland} et~al.}}]{2016arXiv161100036D}
\bibinfo{author}{\bibnamefont{{DESI Collaboration}}},
  \bibinfo{author}{\bibfnamefont{A.}~\bibnamefont{{Aghamousa}}},
  \bibinfo{author}{\bibfnamefont{J.}~\bibnamefont{{Aguilar}}},
  \bibinfo{author}{\bibfnamefont{S.}~\bibnamefont{{Ahlen}}},
  \bibinfo{author}{\bibfnamefont{S.}~\bibnamefont{{Alam}}},
  \bibinfo{author}{\bibfnamefont{L.~E.} \bibnamefont{{Allen}}},
  \bibinfo{author}{\bibfnamefont{C.}~\bibnamefont{{Allende Prieto}}},
  \bibinfo{author}{\bibfnamefont{J.}~\bibnamefont{{Annis}}},
  \bibinfo{author}{\bibfnamefont{S.}~\bibnamefont{{Bailey}}},
  \bibinfo{author}{\bibfnamefont{C.}~\bibnamefont{{Balland}}},
  \bibnamefont{et~al.}, \bibinfo{journal}{ArXiv e-prints}
  (\bibinfo{year}{2016}), \eprint{1611.00036}.

\bibitem[{\citenamefont{{Richard} et~al.}(2019)\citenamefont{{Richard},
  {Kneib}, {Blake}, {Raichoor}, {Comparat}, {Shanks}, {Sorce}, {Sahl{\'e}n},
  {Howlett}, {Tempel} et~al.}}]{2019Msngr.175...50R}
\bibinfo{author}{\bibfnamefont{J.}~\bibnamefont{{Richard}}},
  \bibinfo{author}{\bibfnamefont{J.~P.} \bibnamefont{{Kneib}}},
  \bibinfo{author}{\bibfnamefont{C.}~\bibnamefont{{Blake}}},
  \bibinfo{author}{\bibfnamefont{A.}~\bibnamefont{{Raichoor}}},
  \bibinfo{author}{\bibfnamefont{J.}~\bibnamefont{{Comparat}}},
  \bibinfo{author}{\bibfnamefont{T.}~\bibnamefont{{Shanks}}},
  \bibinfo{author}{\bibfnamefont{J.}~\bibnamefont{{Sorce}}},
  \bibinfo{author}{\bibfnamefont{M.}~\bibnamefont{{Sahl{\'e}n}}},
  \bibinfo{author}{\bibfnamefont{C.}~\bibnamefont{{Howlett}}},
  \bibinfo{author}{\bibfnamefont{E.}~\bibnamefont{{Tempel}}},
  \bibnamefont{et~al.}, \bibinfo{journal}{The Messenger}
  \textbf{\bibinfo{volume}{175}}, \bibinfo{pages}{50} (\bibinfo{year}{2019}),
  \eprint{1903.02474}.

\bibitem[{\citenamefont{{Wang} et~al.}(2021)\citenamefont{{Wang}, {Zhai},
  {Alavi}, {Massara}, {Pisani}, {Benson}, {Hirata}, {Samushia}, {Weinberg},
  {Colbert} et~al.}}]{2021arXiv211001829W}
\bibinfo{author}{\bibfnamefont{Y.}~\bibnamefont{{Wang}}},
  \bibinfo{author}{\bibfnamefont{Z.}~\bibnamefont{{Zhai}}},
  \bibinfo{author}{\bibfnamefont{A.}~\bibnamefont{{Alavi}}},
  \bibinfo{author}{\bibfnamefont{E.}~\bibnamefont{{Massara}}},
  \bibinfo{author}{\bibfnamefont{A.}~\bibnamefont{{Pisani}}},
  \bibinfo{author}{\bibfnamefont{A.}~\bibnamefont{{Benson}}},
  \bibinfo{author}{\bibfnamefont{C.~M.} \bibnamefont{{Hirata}}},
  \bibinfo{author}{\bibfnamefont{L.}~\bibnamefont{{Samushia}}},
  \bibinfo{author}{\bibfnamefont{D.~H.} \bibnamefont{{Weinberg}}},
  \bibinfo{author}{\bibfnamefont{J.}~\bibnamefont{{Colbert}}},
  \bibnamefont{et~al.}, \bibinfo{journal}{arXiv e-prints}
  \bibinfo{eid}{arXiv:2110.01829} (\bibinfo{year}{2021}), \eprint{2110.01829}.

\bibitem[{\citenamefont{{Takada} et~al.}(2014)\citenamefont{{Takada}, {Ellis},
  {Chiba}, {Greene}, {Aihara}, {Arimoto}, {Bundy}, {Cohen}, {Dor{\'e}},
  {Graves} et~al.}}]{2014PASJ...66R...1T}
\bibinfo{author}{\bibfnamefont{M.}~\bibnamefont{{Takada}}},
  \bibinfo{author}{\bibfnamefont{R.~S.} \bibnamefont{{Ellis}}},
  \bibinfo{author}{\bibfnamefont{M.}~\bibnamefont{{Chiba}}},
  \bibinfo{author}{\bibfnamefont{J.~E.} \bibnamefont{{Greene}}},
  \bibinfo{author}{\bibfnamefont{H.}~\bibnamefont{{Aihara}}},
  \bibinfo{author}{\bibfnamefont{N.}~\bibnamefont{{Arimoto}}},
  \bibinfo{author}{\bibfnamefont{K.}~\bibnamefont{{Bundy}}},
  \bibinfo{author}{\bibfnamefont{J.}~\bibnamefont{{Cohen}}},
  \bibinfo{author}{\bibfnamefont{O.}~\bibnamefont{{Dor{\'e}}}},
  \bibinfo{author}{\bibfnamefont{G.}~\bibnamefont{{Graves}}},
  \bibnamefont{et~al.}, \bibinfo{journal}{\pasj} \textbf{\bibinfo{volume}{66}},
  \bibinfo{eid}{R1} (\bibinfo{year}{2014}), \eprint{1206.0737}.

\bibitem[{\citenamefont{{Addison} et~al.}(2013)\citenamefont{{Addison},
  {Hinshaw}, and {Halpern}}}]{2013MNRAS.436.1674A}
\bibinfo{author}{\bibfnamefont{G.~E.} \bibnamefont{{Addison}}},
  \bibinfo{author}{\bibfnamefont{G.}~\bibnamefont{{Hinshaw}}},
  \bibnamefont{and}
  \bibinfo{author}{\bibfnamefont{M.}~\bibnamefont{{Halpern}}},
  \bibinfo{journal}{\mnras} \textbf{\bibinfo{volume}{436}},
  \bibinfo{pages}{1674} (\bibinfo{year}{2013}), \eprint{1304.6984}.

\bibitem[{\citenamefont{{Aubourg} et~al.}(2015)\citenamefont{{Aubourg},
  {Bailey}, {Bautista}, {Beutler}, {Bhardwaj}, {Bizyaev}, {Blanton},
  {Blomqvist}, {Bolton}, {Bovy} et~al.}}]{2015PhRvD..92l3516A}
\bibinfo{author}{\bibfnamefont{{\'E}.}~\bibnamefont{{Aubourg}}},
  \bibinfo{author}{\bibfnamefont{S.}~\bibnamefont{{Bailey}}},
  \bibinfo{author}{\bibfnamefont{J.~E.} \bibnamefont{{Bautista}}},
  \bibinfo{author}{\bibfnamefont{F.}~\bibnamefont{{Beutler}}},
  \bibinfo{author}{\bibfnamefont{V.}~\bibnamefont{{Bhardwaj}}},
  \bibinfo{author}{\bibfnamefont{D.}~\bibnamefont{{Bizyaev}}},
  \bibinfo{author}{\bibfnamefont{M.}~\bibnamefont{{Blanton}}},
  \bibinfo{author}{\bibfnamefont{M.}~\bibnamefont{{Blomqvist}}},
  \bibinfo{author}{\bibfnamefont{A.~S.} \bibnamefont{{Bolton}}},
  \bibinfo{author}{\bibfnamefont{J.}~\bibnamefont{{Bovy}}},
  \bibnamefont{et~al.}, \bibinfo{journal}{\prd} \textbf{\bibinfo{volume}{92}},
  \bibinfo{eid}{123516} (\bibinfo{year}{2015}), \eprint{1411.1074}.

\bibitem[{\citenamefont{{Nadathur} et~al.}(2020)\citenamefont{{Nadathur},
  {Percival}, {Beutler}, and {Winther}}}]{2020PhRvL.124v1301N}
\bibinfo{author}{\bibfnamefont{S.}~\bibnamefont{{Nadathur}}},
  \bibinfo{author}{\bibfnamefont{W.~J.} \bibnamefont{{Percival}}},
  \bibinfo{author}{\bibfnamefont{F.}~\bibnamefont{{Beutler}}},
  \bibnamefont{and} \bibinfo{author}{\bibfnamefont{H.~A.}
  \bibnamefont{{Winther}}}, \bibinfo{journal}{\prl}
  \textbf{\bibinfo{volume}{124}}, \bibinfo{eid}{221301} (\bibinfo{year}{2020}),
  \eprint{2001.11044}.

\bibitem[{\citenamefont{{Alam} et~al.}(2021)\citenamefont{{Alam}, {Aubert},
  {Avila}, {Balland}, {Bautista}, {Bershady}, {Bizyaev}, {Blanton}, {Bolton},
  {Bovy} et~al.}}]{2021PhRvD.103h3533A}
\bibinfo{author}{\bibfnamefont{S.}~\bibnamefont{{Alam}}},
  \bibinfo{author}{\bibfnamefont{M.}~\bibnamefont{{Aubert}}},
  \bibinfo{author}{\bibfnamefont{S.}~\bibnamefont{{Avila}}},
  \bibinfo{author}{\bibfnamefont{C.}~\bibnamefont{{Balland}}},
  \bibinfo{author}{\bibfnamefont{J.~E.} \bibnamefont{{Bautista}}},
  \bibinfo{author}{\bibfnamefont{M.~A.} \bibnamefont{{Bershady}}},
  \bibinfo{author}{\bibfnamefont{D.}~\bibnamefont{{Bizyaev}}},
  \bibinfo{author}{\bibfnamefont{M.~R.} \bibnamefont{{Blanton}}},
  \bibinfo{author}{\bibfnamefont{A.~S.} \bibnamefont{{Bolton}}},
  \bibinfo{author}{\bibfnamefont{J.}~\bibnamefont{{Bovy}}},
  \bibnamefont{et~al.}, \bibinfo{journal}{\prd} \textbf{\bibinfo{volume}{103}},
  \bibinfo{eid}{083533} (\bibinfo{year}{2021}), \eprint{2007.08991}.

\bibitem[{\citenamefont{{Blot} et~al.}(2021)\citenamefont{{Blot}, {Corasaniti},
  {Rasera}, and {Agarwal}}}]{2021MNRAS.500.2532B}
\bibinfo{author}{\bibfnamefont{L.}~\bibnamefont{{Blot}}},
  \bibinfo{author}{\bibfnamefont{P.-S.} \bibnamefont{{Corasaniti}}},
  \bibinfo{author}{\bibfnamefont{Y.}~\bibnamefont{{Rasera}}}, \bibnamefont{and}
  \bibinfo{author}{\bibfnamefont{S.}~\bibnamefont{{Agarwal}}},
  \bibinfo{journal}{\mnras} \textbf{\bibinfo{volume}{500}},
  \bibinfo{pages}{2532} (\bibinfo{year}{2021}), \eprint{2007.14984}.

\bibitem[{\citenamefont{{Efstathiou} and
  {Gratton}}(2020)}]{2020MNRAS.496L..91E}
\bibinfo{author}{\bibfnamefont{G.}~\bibnamefont{{Efstathiou}}}
  \bibnamefont{and}
  \bibinfo{author}{\bibfnamefont{S.}~\bibnamefont{{Gratton}}},
  \bibinfo{journal}{\mnras} \textbf{\bibinfo{volume}{496}},
  \bibinfo{pages}{L91} (\bibinfo{year}{2020}), \eprint{2002.06892}.

\bibitem[{\citenamefont{{Perivolaropoulos} and
  {Skara}}(2021)}]{2021arXiv210505208P}
\bibinfo{author}{\bibfnamefont{L.}~\bibnamefont{{Perivolaropoulos}}}
  \bibnamefont{and} \bibinfo{author}{\bibfnamefont{F.}~\bibnamefont{{Skara}}},
  \bibinfo{journal}{arXiv e-prints} \bibinfo{eid}{arXiv:2105.05208}
  (\bibinfo{year}{2021}), \eprint{2105.05208}.

\bibitem[{\citenamefont{{Schwarz} et~al.}(2016)\citenamefont{{Schwarz}, {Copi},
  {Huterer}, and {Starkman}}}]{2016CQGra..33r4001S}
\bibinfo{author}{\bibfnamefont{D.~J.} \bibnamefont{{Schwarz}}},
  \bibinfo{author}{\bibfnamefont{C.~J.} \bibnamefont{{Copi}}},
  \bibinfo{author}{\bibfnamefont{D.}~\bibnamefont{{Huterer}}},
  \bibnamefont{and} \bibinfo{author}{\bibfnamefont{G.~D.}
  \bibnamefont{{Starkman}}}, \bibinfo{journal}{Classical and Quantum Gravity}
  \textbf{\bibinfo{volume}{33}}, \bibinfo{eid}{184001} (\bibinfo{year}{2016}),
  \eprint{1510.07929}.

\bibitem[{\citenamefont{{Planck Collaboration}
  et~al.}(2014)\citenamefont{{Planck Collaboration}, {Ade}, {Aghanim},
  {Armitage-Caplan}, {Arnaud}, {Ashdown}, {Atrio-Barandela}, {Aumont},
  {Baccigalupi}, {Banday} et~al.}}]{2014A&A...571A..16P}
\bibinfo{author}{\bibnamefont{{Planck Collaboration}}},
  \bibinfo{author}{\bibfnamefont{P.~A.~R.} \bibnamefont{{Ade}}},
  \bibinfo{author}{\bibfnamefont{N.}~\bibnamefont{{Aghanim}}},
  \bibinfo{author}{\bibfnamefont{C.}~\bibnamefont{{Armitage-Caplan}}},
  \bibinfo{author}{\bibfnamefont{M.}~\bibnamefont{{Arnaud}}},
  \bibinfo{author}{\bibfnamefont{M.}~\bibnamefont{{Ashdown}}},
  \bibinfo{author}{\bibfnamefont{F.}~\bibnamefont{{Atrio-Barandela}}},
  \bibinfo{author}{\bibfnamefont{J.}~\bibnamefont{{Aumont}}},
  \bibinfo{author}{\bibfnamefont{C.}~\bibnamefont{{Baccigalupi}}},
  \bibinfo{author}{\bibfnamefont{A.~J.} \bibnamefont{{Banday}}},
  \bibnamefont{et~al.}, \bibinfo{journal}{\aap} \textbf{\bibinfo{volume}{571}},
  \bibinfo{eid}{A16} (\bibinfo{year}{2014}), \eprint{1303.5076}.

\end{thebibliography}

\end{document}